\documentclass[a4paper,12pt, epsfig]{article}
\def\letter{0}\def\pr{0}
\pdfoutput=1 %per jhep
\usepackage{epsfig}
\usepackage{epstopdf}
\usepackage{graphicx}
\usepackage{ifthen}
\usepackage{ulem}

\usepackage{amsmath}
\usepackage[psamsfonts]{amssymb}
\usepackage{euscript}

\usepackage{latexsym}
\usepackage[arrow,matrix,curve]{xy}

\usepackage[hypertexnames=false]{hyperref}
\hypersetup{
    colorlinks=true,
    linkcolor=blue,
    filecolor=magenta,   
    citecolor=black,   
    urlcolor=cyan,
    }

\newskip\humongous \humongous=0pt plus 1000pt minus 1000pt

\newif\ifdtup

\def\,{\hspace{-.1cm}}
\def\hsp{,\hspace{.7cm}}

\def\fc#1#2 {\frac{n}{q}#1\frac{n}{q}#2}

\def\kt{\kappa}
\def\kt{\mathfrak{K}}
\def\ks{|\kt\rangle}

\def\rv{{\rm{vac}}}

\newcommand{\vac}{\ensuremath{|0\rangle}}

\renewcommand{\sin}{\textrm{sin}}

\renewcommand{\sinh}{\textrm{sinh}}

\renewcommand{\tanh}{\textrm{tanh}}
\newcommand{\sech}{\textrm{sech}}
\newcommand{\csch}{\textrm{csch}}

\def\sign#1{{\rm sign}\left(#1\right)}

\renewcommand{\theequation}{\arabic{section}.\arabic{equation}}
\renewcommand{\(}{\begin{equation}}
\renewcommand{\)}{end{equation} \vspace{-.05in}\linebreak}

\newcounter{saveeqn}
\newcounter{savealpheqn}

\newcommand{\alpheqn}{\setcounter{saveeqn}{\value{equation}}%
  \stepcounter{saveeqn}\setcounter{equation}{0}%
  \renewcommand{\theequation}{\mbox{\arabic{section}.\arabic{saveeqn}
\alph{equation}}}
  \renewcommand{\)}{\end{equation}}}
\def\part#1{\frac{\partial}{\partial{#1}}}%
\def\group#1{\refstepcounter{equation}\setcounter{saveeqn}
 {\value{equation}}%
  \label{#1}\setcounter{equation}{0}%
\renewcommand{\theequation}{\mbox{\arabic{section}.\arabic{saveeqn}
\alph{equation}}}
  \renewcommand{\)}{\end{equation}}}
\newcommand{\reseteqn}{\setcounter{equation}{\value{saveeqn}}%
  \renewcommand{\theequation}{\arabic{section}.\arabic{equation}}%
  \renewcommand{\)}{\end{equation}}}

\newcommand{\aalpheqn}{\setcounter{saveeqn}{\value{equation}}%
  \stepcounter{saveeqn}\setcounter{equation}{0}%
  \renewcommand{\theequation}{\mbox{
        \Alph{subsection}.\arabic{saveeqn}\alph{equation}}}
   \renewcommand{\)}{\end{equation}}}
\newcommand{\areseteqn}{\setcounter{equation}{\value{saveeqn}}%
  \renewcommand{\theequation}{\Alph{subsection}.\arabic{equation}}%
  \renewcommand{\)}{\end{equation}}}

\renewcommand{\thefootnote}{\alph{footnote}}
\renewcommand{\(}{\begin{equation}}
\renewcommand{\)}{\end{equation}}
\newcommand{\ba}{\begin{eqnarray}}
\newcommand{\ea}{\end{eqnarray}}

\renewcommand{\b}{\beta}

\renewcommand{\sl}{{\sqrt{\lambda}}}

\newcommand{\cbp}{\mathop{\vtop{\ialign{##\crcr
   $\hfil\displaystyle{}\hfil$\crcr\noalign{\kern-13pt\nointerlineskip}
   \BIG{)}\hskip0pt\crcr\noalign{\kern3pt}}}}}
\newcommand{\pa}{\mathop{\vtop{\ialign{##\crcr

$\hfil\displaystyle{\oplus}\hfil$\crcr\noalign{\kern+1pt\nointerlineskip
}
   \hspace{.08in}$^{\alpha=0}$\hskip6pt\crcr\noalign{\kern3pt}}}}}
\renewcommand{\hsp}{,\hspace{.3in}}
\newcommand{\p}{^\prime}

\catcode`\@=11
\def\vereq#1#2{\lower3pt\vbox{\baselineskip1.5pt \lineskip1.5pt
\ialign{$\m@th#1\hfill##\hfil$\crcr#2\crcr\sim\crcr}}}
\catcode`\@=12

\renewcommand{\(}{\begin{equation}}
\renewcommand{\)}{\end{equation}}

\def\pin#1{\int \frac{d#1}{2\pi}}
\def\ppin#1{\int\hspace{-17pt}\sum \frac{d#1}{2\pi}}
\def\ppink#1{\int\hspace{-17pt}\sum\frac{d^{#1}k}{(2\pi)^{#1}}}
\def\ppinkp#1{\int\hspace{-17pt}\sum\frac{d^{#1}k\p}{(2\pi)^{#1}}}
\def\dint{\int\hspace{-12pt}\sum }
\def\pink#1{\int \frac{d^{#1}k}{(2\pi)^{#1}}}

\def\Bd#1{B^\ddag_{k_{#1}}}

\def\cc{\mathcal{C}}
\def\df{\mathcal{D}_{f}}

\def\I{\mathcal{I}}

\def\os{\omega_S}

\def\blu#1{\textcolor{blue}{Jarah: #1}}
\def\red#1{\textcolor{red}{Hui: #1}}

\newcommand{\beas}{\begin{eqnarray*}}
\newcommand{\eeas}{\end{eqnarray*}}

\newcommand{\bquo}{\begin{quote}}
\newcommand{\enqu}{\end{quote}}

\def\lim#1{\stackrel{\rm{lim}}{{}_{#1}}}

\newcommand{\R}{{\mathbb R}}

\newcommand{\g}{\mathfrak g}

\def\ok#1{\omega_{k_{#1}}}
\def\okp#1{\omega_{k\p_{#1}}}
\def\okt#1{\omega_{\kt_{#1}}}
\def\V#1{V^{(#1)}(\sqrt{\lambda}f(x))}
\def\v#1{V^{(#1)}[f(x),x]}
\def\ck{\csch\left(\frac{\pi k}{2\b}\right)}

\def\mb{\mathcal{B}}
\def\mc{\mathcal{C}}
\def\md{\mathcal{D}}
\def\me{\mathcal{E}}

\newcommand{\beq}{\begin{equation}}
\newcommand{\eeq}{\end{equation}}
\newcommand{\bea}{\begin{eqnarray}}
\newcommand{\eea}{\end{eqnarray}}

\newskip\humongous \humongous=0pt plus 1000pt minus 1000pt

\newif\ifdtup

\catcode`\@=11

\ifthenelse{\equal{\letter}{0}}{ %se lettere=0, commenta questo

\@addtoreset{equation}{section}
\def\theequation{\arabic{section}.\arabic{equation}}

\def\@normalsize{\@setsize\normalsize{15pt}\xiipt\@xiipt
\abovedisplayskip 14pt plus3pt minus3pt%
\belowdisplayskip \abovedisplayskip
\abovedisplayshortskip \z@ plus3pt%
\belowdisplayshortskip 7pt plus3.5pt minus0pt}

\def\small{\@setsize\small{13.6pt}\xipt\@xipt
\abovedisplayskip 13pt plus3pt minus3pt%
\belowdisplayskip \abovedisplayskip
\abovedisplayshortskip \z@ plus3pt%
\belowdisplayshortskip 7pt plus3.5pt minus0pt
\def\@listi{\parsep 4.5pt plus 2pt minus 1pt
      \itemsep \parsep
      \topsep 9pt plus 3pt minus 3pt}}

\relax

\def\section{\@startsection{section}{1}{\z@}{3.5ex plus 1ex minus  .2ex}{2.3ex plus .2ex}{\large\bf}}

\def\thesection{\arabic{section}}
\def\thesubsection{\arabic{section}.\arabic{subsection}}

\def\appendix{\setcounter{section}{0}
 \def\thesection{Appendix \Alph{section}}
 \def\thesubsection{\Alph{section}.\arabic{subsection}}
 \def\theequation{\Alph{section}.\arabic{equation}}}
\renewcommand{\theequation}{\arabic{section}.\arabic{equation}}

}{
\renewcommand{\theequation}{\arabic{equation}}

} %se letter=0, commenta questo

\begin{document}
% ========================================================================
\def\thefootnote{\fnsymbol{footnote}}
\def\thetitle{A Reduced Inner Product for Kink States}
\def\auttwo{Hui Liu}
\def\autone{Jarah Evslin}
\def\affc{School of Fundamental Physics and Mathematical Sciences, Hangzhou Institute for Advanced Study,
University of Chinese Academy of Sciences, Hangzhou 310024, China}
\def\affb{University of the Chinese Academy of Sciences, YuQuanLu 19A, Beijing 100049, China}
\def\affd{Arnold Sommerfeld Center, Ludwig-Maximilians-Universität, Theresienstraße 37, 80333 München, Germany}
\def\affa{Institute of Modern Physics, NanChangLu 509, Lanzhou 730000, China}
\def\affe{Institute of Contemporary Mathematics, School of Mathematics and Statistics,
Henan University, Kaifeng, Henan 475004, P. R. China}
%\def\affc{Departamento de F\'isica de Part\'iculas, Universidad deSantiago de Compostela and Instituto Galego de F\'isica de Altas Enerxias% (IGFAE), E-15782
%Santiago de Compostela, Spain}
%\def\affd{Physics Dept., Brookhaven National Laboratory, Bldg. 510A, Upton, NY 11973, USA}
%\def\affc{Wigner Research Centre, H-1525 Budapest 114, P.O.B. 49, Hungary}

%\title{Titolo}

\ifthenelse{\equal{\pr}{1}}{
\title{\thetitle}
\author{\autone}
\author{\auttwo}
\author{\autthree}
\affiliation {\affa}
\affiliation {\affb}
%\affiliation {\affc}
%\affiliation {\affd}
%pr e uno

}{

\begin{center}
{\large {\bf \thetitle}}

\bigskip

\bigskip

%\catcode`@=11

{\large \noindent  \autone{${}^{1,2}$} \footnote{jarah@impcas.ac.cn} %{
and \auttwo{${}^{3,2,4}$} \footnote{hui.liu@campus.lmu.de}
%and \autthree{${}^{5}$} \footnote{byzhang@henu.edu.cn}
}

%{\large \noindent  \autone{${}^{1,2}$} \footnote{jarah@impcas.ac.cn} and \auttwo{${}^{1,2}$} \footnote{guohengyuan@impcas.ac.cn}}

\vskip.7cm

1) \affa\\
2) \affb\\
3) \affc\\
4) \affd\\
%5) \affe\\

\end{center}

}

\begin{abstract}
\noindent
Solitons in classical field theories correspond to states in quantum field theories.  If the spatial dimension is infinite, then momentum eigenstates are not normalizable.  This leads to infrared divergences, which are generally regularized via wave packets or by compactification.  However, in some applications both possibilities are undesirable.  In the present note, we introduce a finite inner product on translation-invariant kink states that allows us to compute probabilities involving these nonnormalizable states.  Essentially, it is the quotient of the usual inner product by the translation group. We present a surprisingly simple formula for the reduced inner product, which requires no knowledge of the zero-mode dependence of the states but includes a correction which accounts for the mixing between zero modes and normal modes as the kink moves.  As an application, we show that initial and final state corrections to meson multiplication vanish.  However, we find that the pole of the subleading term in the initial state requires an infinitesimal imaginary shift.

%In a (1+1)-dimensional scalar quantum field theory, we calculate the leading-order probability of meson multiplication, which is the inelastic scattering process: kink + meson $\rightarrow$ kink + 2 mesons.  We also calculate the differential probability with respect to the final meson momenta and the probability that one or two of the final mesons recoils back towards the source.  In the ultrarelativistic limit of the initial meson, the total probability tends to a constant, which we calculate analytically in the $\phi^4$ model.  At this order the meson sector conserves energy on its own, while the incoming meson applies a positive pressure to the kink.  This is in contrast with the situation in classical field theory, where Romanczukiewicz and collaborators have shown that, in the presence of a reflectionless kink, only meson fusion is allowed, resulting in a negative radiation pressure on the kink.

\end{abstract}

% \vfill
%
% \end{titlepage}
\setcounter{footnote}{0}
\renewcommand{\thefootnote}{\arabic{footnote}}

\ifthenelse{\equal{\pr}{1}}
{
\maketitle
}{}

\section{Introduction}

\subsection{Motivation}

The collective coordinate method of Ref.~\cite{gjscc} allows for arbitrary calculations involving quantum kinks\footnote{At one loop, there are many robust and efficient methods for treating solitons, beginning with Ref.~\cite{dhn2}.  Ref.~\cite{wrev} provides a recent review.}.  The position of the kink itself is quantized, and the fields are expanded about this time-dependent position.  However, the interplay of the kink position and the field expansion is complicated.  To bring the operators into a canonical form, to allow for quantization, one requires a nonlinear canonical transformation already in the classical theory.  This transformation does not leave the quantum path integral invariant, and so in the quantum theory, an infinite series of terms needs to be added to the Hamiltonian \cite{gj76}.  These complications have made all but the simplest problems impractical.  For example, two-loop corrections to kink masses have only been computed when they are already known as a result of integrabilility \cite{vega,verwaest} or supersymmetry \cite{shifmananomalia}.  Also, kink-meson scattering has been restricted to calculating the leading contribution to an effective Yukawa coupling \cite{hayashi1,hayashi2}.  However, recently it is led to promising developments in the calculation of form factors \cite{accel,andyprl,raggio}.

A new, simpler method, linearized kink perturbation theory, has been formulated in Refs.~\cite{mekink,me2loop}.  Here the kink fields are expanded as if the kink were at a fixed base point.   As a result, the fields are canonical from the beginning.  The price is that the distance of the kink from the base point is treated perturbatively, as a semiclassical expansion in the coupling.  Thus it is not reliable if the kink wave packet extends beyond the radius of convergence of the expansion.  The radius of convergence, in the sense of an asymptotic series, is more than the de Broglie wavelength of the kink, but less than its classical diameter.  This leaves the method applicable to localized kink wave packets\footnote{One should draw a distinction between a wave packet for the center of mass of the kink-meson system, whose size is treated perturbatively and thus is bounded, and a wave packet describing the relative position of a meson with respect to the kink, which is treated exactly and whose monochromatic limit poses no complications.}, or more generally localized soliton wave packets, as arise in many applications such as solitonic dark matter \cite{memono,fuzzy14}, pinned Abrikosov vortices and kink-impurity interactions \cite{impure,chris}.

However, sometimes one is interested in the opposite regime, in which the kink is in a translation-invariant state, such as its ground state or the ground state of a system of a kink and a finite number of mesons.  A translation-invariant state is a quantum superposition, summing over all possible simultaneous and equal translations of the kink and mesons, which necessarily keep the relative distances fixed.  This is relevant \cite{hayashirep}, for example, to treating proton-meson scattering using the Skyrme \cite{skyrme,smorg} model.  Here one must simultaneously consider kinks at positions arbitrarily far from the base point.  The perturbative approach above naively fails miserably.

The solution to this problem is to use translation-invariance\footnote{An alternative approach, which does not require translation-invariance, was presented in Ref.~\cite{point22}.  However so far zero-modes have not been included.}.  All of the information regarding a translation-invariant kink state is contained in the configuration involving the kink at the base point, and so a study of that case, together with translation-invariance, yields all quantities.  In the case of computations of kink masses, this is achieved \cite{me2loop} by projecting the state, perturbatively, onto the kernel of the momentum operator and then solving the Schrodinger equation for the power series expansion in the kink position about the base point.  If it is solved for all coefficients in the expansion about the base point, it is solved everywhere by translation invariance.  And indeed, the method has been shown to agree with previous calculations of form factors and two-loop mass corrections where available and, due to its simplicity, it has provided novel calculations of the mass of non-integrable, non-supersymmetric kinks \cite{phi42loop} and even excited kinks \cite{menormal}, as well as kink form factors in non-integrable models \cite{hengyuan}.

However, this problem becomes more severe when one tries to compute dynamical quantities.  Here one needs to calculate inner products of states.  These translation-invariant states are non-normalizable, and so their inner products do not exist.  Usually one can evade this problem by regularizing the state in the form of a wave packet and taking the limit in which the wave packet becomes large.  Unfortunately, in the case of linearized perturbation theory, this cannot be done as there is no way to treat a finite wave packet which is larger than the radius of convergence.  One may try to avoid the problem by compactifying space.  However, such a compactification requires particular boundary conditions and there is no guarantee that finite contributions do not remain when the compactification radius is taken to infinity.  Thus, if compactification can be avoided, it is better in our opinion to avoid it.

So far in dynamical problems we have side-stepped this complication.  In the case of excited kink decay \cite{alberto} and meson multiplication \cite{memult} the inner products that appeared were always in fractions where the same inner product appeared in both the numerator and the denominator of probabilities, and so we canceled them.  However, at higher orders, different states will appear in the numerator and denominator.  %Also, in the case of meson multiplication, we ignored corrections to probabilities arising from quantum corrections to initial and final states, although in principle these were of the same order as the amplitude which was calculated.  

\subsection{Reduced Inner Product}

In this note we suggest a new strategy for dealing with norms of translation-invariant states without regularizing the infinities.  As the inner products of interest always appear in both the numerator and denominator of an expression for an observable, we quotient both by the infinite volume translation group, being careful to keep the relevant Jacobian factor.  This strategy resembles gauging by the global translation symmetry, which simultaneously shifts the kink and also the mesons.  Intuitively this is also related to a compactification of radius zero, except that the distance between the kink and mesons is preserved and so they are effectively in an infinite space.

We restrict our attention to the kink sector.  This is the Fock space of any finite number of mesons in the presence of a quantum kink.  However a generalization to other topological sectors is obvious.

Our main result, a formula for the reduced inner product of any two translation-invariant states, is presented in Eq.~(\ref{padqft}).   Intuitively, the ordinary inner product can be written as an integral over the collective coordinate $x$ and the reduced inner product is defined by inserting $\delta(x)$.  The collective coordinate transforms under translations via the usual rigid shift.  In linearized perturbation theory one works using not the collective coordinate $x$, but rather the linearized coordinate $y$, whose transformation under translations is rather complicated.  Our formula (\ref{padqft}) replaces the $\delta(x)$ with $y\p(x)\delta(y)$, where the Jacobian factor $y\p(x)$ is an operator. Amazingly, as a result of the $\delta(y)$, this formula is simpler than the usual formula for the inner product, as it does not use the dependence of the state on the zero-modes, which have eigenvalue $y$.  This is not obviously inconsistent,  as, in the case of translation-invariant states, the zero-mode dependence is entirely fixed by the translation invariance \cite{me2loop}.  The price for this simplification is the addition $y\p(x)$, which contains two finite quantum corrections above the naive inner product, which mix sectors whose meson numbers differ by one unit.  These corrections reflect the fact that the kink zero-mode mixes with the normal modes upon translation.

%The idea proceeds in a few steps.  First, one constructs an orthogonal basis of translation-invariant states for this sector.  For each state in the basis, one then calculates a reduced norm.  The reduced norm is the key to our construction.  The naive norm would be infinite.  The reduced norm is the norm of the state quotiented by the translation-invariance.  One can check that this basis is indeed orthogonal in terms of the reduced norm and if not, modify the basis.  Next, when one encounters an inner product, one decomposes each state in terms of the basis and uses the reduced norms of the basis and linearity to evaluate the reduced inner product.  When calculating probabilities, one always finds the same number of inner products in the numerator as in the denominator, and so one simply replaces the true inner products with the reduced inner products, leaving volumes of the translation group which cancel in the numerator and denominator.

Three applications are presented.  First, this allows us to place formal manipulations, in which these norms were canceled in Refs.~\cite{alberto,memult}, on more solid footing.  Also, it allows us to treat cases where more complicated inner products arise, such as matrix elements of zero-modes.  In fact, this already happens in the order $O(\sqrt{\lambda})$ contribution to the meson multiplication amplitude, arising from an $O(\sqrt{\lambda})$ correction to the initial or final state and no interaction.  We thus apply our formalism to calculate these corrections, and to show that they vanish at this order.  

Finally, we use this to calculate the leading order correction to the one-meson state consisting of two-meson states.  It was already calculated in Ref.~\cite{menormal} but the result involved a pole at a location where the Hamiltonian is degenerate, and so distinct prescriptions for treating the pole yield legitimate, yet inequivalent, Hamiltonian eigenstates.  We find that one particular prescription for the pole yields the physically-motivated initial conditions for meson-kink scattering, in which the initial state never contains two mesons.

In Sec.~\ref{revsez} we review the linearized kink perturbation theory of Refs.~\cite{mekink} and \cite{me2loop}.  Next in Sec.~\ref{qmsez} we present our construction of the reduced inner product in quantum mechanics.  This construction is adapted to kink sectors of quantum field theories in Sec.~\ref{qftsez}.  In Sec.~\ref{exsez} we provide some examples of reduced inner products.  Finally in Sec.~\ref{multsez} we apply this formalism to evaluate the meson multiplication amplitude using translation-invariant states, finding the same result as Ref.~\cite{memult} where quantum corrections to the initial and states were ignored.  In addition, we find the quantum corrections to the initial state which are relevant to this experiment, corresponding to a prescription for treating the pole in Ref.~\cite{menormal}.

\section{Review} \label{revsez}

While we suspect that it may be generalized to theories of greater phenomenological interest, so far linearized kink perturbation theory has only been formulated for (1+1)-dimensional quantum field theories with a Schrodinger picture scalar field $\phi(x)$, conjugate to $\pi(x)$, and a Hamiltonian
\begin{equation}
H=\int d x: \mathcal{H}(x):_a\hsp \mathcal{H}(x)=\frac{\pi^2(x)}{2}+\frac{\left(\partial_x \phi(x)\right)^2}{2}+\frac{V(\sqrt{\lambda} \phi(x))}{\lambda}.
\end{equation}
The potential $V$ has degenerate minima and a classical kink solution $f(x)$ interpolates from one to another.  The normal ordering $::_a$ renders such theories UV-finite.  It is defined at the mass scale $m$, defined by
\beq
m^2=V^{(2)}(\sqrt{\lambda} f(\pm \infty))\hsp
V^{(n)}(\sqrt{\lambda} \phi(x))=\frac{\partial^n V(\sqrt{\lambda} \phi(x))}{(\partial \sqrt{\lambda} \phi(x))^n}.
\eeq
This in fact defines two values of the mass, one at the vacuum on each side of the kink.  If the masses are different, then one-loop corrections to the vacuum energy imply that one vacuum is a false vacuum, and the kink will accelerate towards it \cite{wstabile}.  Such a kink does not correspond to a Hamiltonian eigenstate in the quantum theory and we will not consider it further.  We will treat the theory using a semiclassical expansion in the coupling $\sqrt{\lambda}$.

We will consider several sectors of the Hilbert space.  The vacuum sector consists of configurations with no kinks, and a finite number of perturbative excitations of $\phi(x)$, which we will call mesons.  Here, one meson is a plane wave, which is created by a creation operator defined using the usual plane wave decomposition of $\phi(x)$ and $\pi(x)$.  More precisely, there is a vacuum sector for each minimum of the classical potential $V$, and sometimes one needs to distinguish between the vacuum sectors to the left and to the right of the kink.  %In these instances, we will refer to them as the left and right vacuum sectors.  

The kink sector consists of a single kink and a finite number of excitations.  We will refer to the excitations which are unbound again as mesons, and those which are bound as shape modes.  These are also created by creation operators, $B^\ddag_S$ and $B^\ddag_k$ respectively, defined by the decomposition of $\phi(x)$ and $\pi(x)$ in terms of normal modes $\g(x)$ \cite{cahill76}
\bea
\phi(x) &=&\phi_0 \mathfrak{g}_B(x)+\ppin{k} \left(B_k^{\ddag}+\frac{B_{-k}}{2 \omega_k}\right) \mathfrak{g}_k(x)\hsp
B^\ddag_k=\frac{B^\dag_k}{2\ok{}}\hsp
B^\ddag_S=\frac{B^\dag_S}{2\omega_S} \label{dec}\\
\pi(x) &=&\pi_0 \mathfrak{g}_B(x)+i \ppin{k}\left(\omega_k B_k^{\ddag}-\frac{B_{-k}}{2}\right) \mathfrak{g}_k(x)\hsp
B_S=B_{-S}\hsp \ppin{k}=\pin{k}+\sum_S \nonumber
\eea
where $\phi_0$ is the zero mode.% whose eigenvalue yields the kink's position and $\pi_0$ is the proportional to the kink's momentum.    

The normal modes $\g(x)$ are constant frequency $\omega$ solutions of the Sturm-Liouville equation for infinitesimal perturbations about a kink
\beq
\V{2}{\g}(x)=\omega^2{\g}(x)+{\g}^{\prime\prime}(x)\hsp \phi(x,t)=e^{-i\omega t}\g(x). \label{sl}
\eeq
There will always be one solution, $\g_B(x)$, with $\omega_B=0$ corresponding to a zero mode.  The shape modes are those with $0<\omega_S<m$.  Continuum modes have frequencies 
\beq
\ok{}=\sqrt{m^2+k^2}.
\eeq
All modes are assembled and normalized to satisfy $\g^*_k=\g_{-k}$ and the completeness relations
\beq
\int dx |{\g}_{B}(x)|^2=1,\
\int dx {\g}_{k_1} (x) {\g}^*_{k_2}(x)=2\pi \delta(k_1-k_2),\ 
\int dx {\g}_{S_1}(x){\g}^*_{S_2}(x)=\delta_{S_1S_2}. \label{comp}
\eeq
We fix the sign of $\g_B$ via
\beq
\g_B(x)=-\frac{f\p(x)}{\sqrt{Q_0}} \label{gb}
\eeq
where $Q_i$ is the $i$-loop correction to the energy of the ground state kink.  Note that the sign is not the same as in previous papers.  $Q_0$ is just the energy of the classical field configuration.

Any operator can be expanded in terms of $\phi(x)$ and $\pi(x)$ or alternatively in terms of $\phi_0,\ \pi_0,\ B^\ddag_k,\ B_k,\ B^\ddag_S$\ and\ $B_S$.  From the canonical commutation relations for $\phi(x)$ and $\pi(x)$, we find the algebra satisfied by this second basis
\beq
\left[\phi_0, \pi_0\right]=i, \quad\left[B_{S_1}, B_{S_2}^{\ddagger}\right]=\delta_{S_1 S_2}, \quad\left[B_{k_1}, B_{k_2}^{\ddagger}\right]=2 \pi \delta\left(k_1-k_2\right).
\eeq

We would like to perform calculations involving states in the one-kink sector.  The problem is that these states are nonperturbative.  This is easy to understand in classical field theory, where small perturbations about the kink correspond to field configurations $\phi(x,t)$ close to $f(x)$, which is far from zero, and so higher moments of $\phi(x,t)$ are not small.  The solution in classical field theory is to decompose $\phi(x,t)=f(x)+\eta(x,t)$ and work with $\eta(x,t)$, which is small and so can be treated perturbatively.  We would like an analogous procedure in the quantum theory, where $\phi(x)$ is a Schrodinger picture quantum field.

The replacement $\phi(x,t)\rightarrow\eta(x,t)$ in the classical theory can be achieved, in the quantum theory, by conjugating with the displacement operator $\df$
\beq
\df={{\rm Exp}}\left[-i\int dx f(x)\pi(x)\right]\hsp
\df^\dag \phi(x) \df = \phi(x)-f(x).  \label{dfd}
\eeq
This displacement operator is unitary and commutes with normal ordering.  It also acts on the states, mapping a vacuum sector state to a one-kink sector state.

So far we have worked in the defining frame of the Hilbert space.  This is the usual representation in which the Hamiltonian $H$ generates time translations and the momentum $P$ generates spatial translations.  Energies are eigenvalues of $H$ while $e^{-iHt}$ yields finite time evolution.  All states in all sectors can be written in the defining frame, using Dirac kets.

Now we want to write the same Hilbert space in a new frame, called the kink frame.  We define the state $|\psi\rangle$ in the kink frame to be the state $\df|\psi\rangle$ in the defining frame.  In this definition, $\df$ plays the role of a passive transformation, changing the coordinate system used to describe the Hilbert space without changing the state.  This passive transformation transforms not the states but rather the operators that act on these states.  For example, time and space translations in the kink frame are generated by the kink Hamiltonian $H\p$ and kink momentum $P\p$
\beq
H\p=\df^\dag H\df\hsp
P\p=\df^\dag P\df=P+\sqrt{Q_0}\pi_0. \label{df}
\eeq
As a consistency check, note that the energy of $|\psi\rangle$ in the kink frame, as measured by $H\p$, is equal to the energy of $\df|\psi\rangle$ in the vacuum frame, as measured by $H$
\beq
H\df|\psi\rangle=E\df|\psi\rangle\Rightarrow H\p|\psi\rangle=\df^\dag H\df|\psi\rangle=E|\psi\rangle.
\eeq
This is a trivial manipulation, but for kink states the eigenvalue equation for $H\p$ is perturbative while for $H$ it is nonperturbative.  Thus in the kink frame, kink states are within the range of perturbation theory, just like $\eta(x,t)$ in classical field theory.  Thus one can find kink states perturbatively in the kink frame, and if desired they can be transformed back to the defining frame using $\df$.

We expand the kink Hamiltonian
\beq
H\p=\sum_{i=0}^\infty H\p_i \label{semi}
\eeq
where $H\p_i$ is of order $O(\lambda^{i/2-1})$.  The terms $H\p_i$ were found in Ref.~\cite{mekink}
\bea
H\p_0&=&Q_0\hsp H\p_1=0\hsp
H\p_2=Q_1+H\p_{\text {free }}, \quad H\p_{\text {free }}=\frac{\pi_0^2}{2}+\omega_S B_S^{\ddag} B_S+\int \frac{d k}{2 \pi} \omega_k B_k^{\ddag} B_k
\nonumber\\
H\p_{n>2}&=&\lambda^{\frac{n}{2}-1}\int dx \frac{V^{(n)}(\sqrt{\lambda} f(x))}{n !}: \phi^n(x):_a.
\eea
Note that the terms in $H\p_{\rm{free}}$ correspond to solved systems in quantum mechanics.  The $\pi_0^2$ term is the kinetic energy for a free particle of mass $Q_0$, and so we see that $\phi_0/\sqrt{Q_0}$ and $\sqrt{Q_0}\pi_0$ are the position and momentum of the center of mass of the kink.  The factor of $\sqrt{Q_0}$ relating the kink position to the eigenvalue of $\phi_0$ will appear again later, as the leading term in the reduced norm. The other terms in $H\p_{\text {free }}$ are harmonic oscillators, one for each shape mode and continuum mode.

All Hamiltonian eigenstates $|\psi\rangle$ will be decomposed in a semiclassical expansion
\beq
|\psi\rangle=\sum_{i=0}^\infty|\psi\rangle_i  \label{semi}
\eeq
where $|\psi\rangle_i$ is of order $O(\lambda^{i/2})$.  The leading components $|\psi\rangle_0$ of all states $|\psi\rangle$ solve the leading order eigenvalue equation for $H\p$, and so are defined to be the eigenstates of $H\p_2$.   

In the kink frame, the ground state of the kink sector is $\vac$.  The leading component $\vac_0$ is the ground state of the system defined by each term in $H\p_{\rm{free}}$ and so satisfies
\beq
\pi_0\vac_0=B_k\vac_0=B_S\vac_0=0. \label{v0}
\eeq
Similarly one may define, at leading order, states with one kink and one or two mesons
\beq
|k\rangle_0=B^\ddag_k\vac_0\hsp
|kk\p\rangle_0=B^\ddag_k B^\ddag_{k\p}\vac_0. \label{2m}
\eeq

\section{Reduced Inner Products in Quantum Mechanics} \label{qmsez}

Our main result will be a finite, reduced inner product for translation-invariant states in the one kink sector.  It is derived by quotienting the ordinary inner product by the translation group, keeping careful track of the Jacobian term.  In this section, we will motivate our result by defining a similar reduced inner product in quantum mechanics.

The Jacobian term is nontrivial because the translation operator $P\p$ acts nonlinearly on the linearized coordinates $y$, defined to be the eigenvalues of $\phi_0$.  However, it acts linearly, as a simple shift, on the collective coordinate $x$.  We will derive our result by first computing the, rather trivial, reduced inner product for a state expressed in collective coordinates $x$, and later will derive a matching condition between collective and linearized coordinates $y$ which allows us to define the reduced inner product on a state expressed in terms of linearized coordinates.

\subsection{Collective Coordinates: Definitions}

We begin by defining the collective coordinate description of states in our quantum mechanical Hilbert space.

Let $|e^n\rangle$ be an orthogonal basis of states which are invariant under the translation operator $P\p$.  Each can be decomposed into eigenstates of the collective coordinate operator $\hat x$
\beq
|e^n\rangle=\int dx |nx\rangle_x\hsp
\hat x |n x\rangle_x=x|n x\rangle_x\hsp
[\hat x,P\p]=i
\eeq
where $n$ is an integer quantum number and
\beq
P\p\int dx F(x) |nx\rangle_x=-i\int dx F\p(x) |nx\rangle_x\hsp
{}_x\langle n_1 x_1|n_2x_2\rangle_x=\delta_{n_1 n_2}\delta(x_1-x_2). \label{xdef}
\eeq
Note that the first relation in (\ref{xdef}) implies that $P\p$ acts on this basis like a momentum operator in quantum mechanics
\beq
[\hat x,e^{-ix_2 P\p}]=x_2e^{-ix_2 P\p}\Rightarrow
e^{-ix_2 P\p}|n x_1\rangle_x=|n,x_1+x_2\rangle_x.
\eeq
%where $\hat x$ is the operator whose eigenvalue is the collective coordinate $x$.  

While the $|e^n\rangle$ are not normalizable, we define the reduced inner product by
\beq\label{redee}
\langle e^m|e^n\rangle_{\rm{red}}=\delta_{mn}.
\eeq
Any translation-invariant $|\psi\rangle$ can be expanded
\beq
|\psi\rangle=\sum_n \psi_n|e^n\rangle.
\eeq
Therefore any reduced inner product is
\beq
\langle \phi|\psi\rangle_{\rm{red}}=\sum_n \phi^*_n \psi_n.
\eeq

\subsection{Linearized Coordinates: Inner Product}

Consider another basis of states $|ny\rangle_y$ where $\phi_0$ and $\pi_0$ are Hermitian operators such that
\beq
\phi_0|ny\rangle_y=y|ny\rangle_y\hsp \pi_0\int dy F(y) |ny\rangle_y=-i\int dy F\p(y)|ny\rangle_y.
\eeq
Define the integral quantum number $n$ such that 
\beq
{}_y\langle n_1 y_1|n_2 y_2\rangle_y=\delta_{n_1n_2} G_{n_1}(y_1,y_2) 
\eeq
for some functions $G_n$.

As $\phi_0$ is Hermitian, 
\beq
0={}_y\langle n y_1|(\phi_0-\phi_0)|n y_2\rangle_y=(y_1-y_2){}_y\langle n y_1|n y_2\rangle_y=(y_1-y_2)G_n(y_1,y_2)
\eeq
and so $G_n(y_1,y_2)$ is only nonvanishing if $y_1=y_2.$  Therefore we will write it simply as  $\delta(y_1-y_2)G_n(y_1)$ and
\beq
{}_y\langle n_1 y_1|n_2 y_2\rangle_y=\delta_{n_1n_2} \delta(y_1-y_2)G_{n_1}(y_1).
\eeq
As $\pi_0$ is Hermitian, for any functions $F_i(y)$ with compact support
\bea
0&=&\int dy_1\int dy_2\ {}_y\langle n y_1| F_1^*(y_1) (\pi_0-\pi_0) F_2(y_2)|n y_2\rangle_y\\
&=&\int dy_1\int dy_2\ {}_y\langle n y_1| \left[i F_1^{\prime *}(y_1)  F_2(y_2)+iF_1^{*}(y_1)  F\p_2(y_2)\right]|n y_2\rangle_y\nonumber\\
&=&i\int dy_1\int dy_2\left[F_1^{\prime *}(y_1)  F_2(y_2)+F_1^{*}(y_1)  F\p_2(y_2) \right]G_n(y_1)\delta(y_1-y_2)\nonumber\\
&=&i\int dy \partial_y(F_1^*(y)F_2(y))G_n(y)=-i\int dy F_1^*(y)F_2(y)\partial_y G_n(y).
\eea
As this is true for arbitrary functions with compact support, we find
\beq
\partial_y G_n(y)=0
\eeq
and so we replace $G_n(y)$ with $G_n$ and write
\beq
{}_y\langle n_1 y_1|n_2 y_2\rangle_y=\delta_{n_1n_2} \delta(y_1-y_2)G_{n_1}.
\eeq
Finally, we may renormalize the $|n y\rangle_y$ states by a factor of $1/\sqrt{G_n}$ so that %note that our choice of quantum number $n$ was arbitrary, so long as distinct quantum numbers were orthogonal.  Therefore, this inner product is diagonal in every orthonormal basis.  This means that $G_n$ is independent of $n$.  We will normalize our states so that it is equal to unity
\beq
{}_y\langle n_1 y_1|n_2 y_2\rangle_y=\delta_{n_1n_2} \delta(y_1-y_2). \label{yort}
\eeq

\subsection{Linearized Coordinates: Translations}

Consider the translation-invariant state
\beq\label{transinvar}
|\psi\rangle=\sum_n \int dy\hat\psi_n(y)|ny\rangle_y
\eeq
and assume that the translation operator is of the form
\beq
P\p=A\pi_0+B+C\phi_0 \label{pp}
\eeq
where $A$, $B$ and $C$ are matrices that commute with $\pi_0$ and $\phi_0$.  Note that the hat notation does not mean that $\hat \psi$ is an operator, but merely that it is the coefficient in the $y$ basis.

Acting the translation operator on the invariant state, one finds
\beq
P\p|\psi\rangle=\sum_{mn}\int dy \left[-iA_{mn}\hat \psi_n\p(y)+ B_{mn}\hat \psi_n(y)+ C_{mn}y\hat \psi_n(y)
\right]|my\rangle_y
\eeq
so that for invariant states
\beq
A\hat \psi\p(y)+i(B+yC)\hat \psi(y)=0.
\eeq
In particular, for small $\epsilon$
\beq
\hat\psi(\epsilon)=\hat\psi(0)+\epsilon\hat\psi\p(0)=\hat\psi(0)-i\epsilon A^{-1}B\hat\psi(0).
\eeq

Let $\hat{v}^j$ be an eigenvector of $A_{mn}$ such that
\beq
\sum_n A_{mn}\hat v^j_n=\lambda_j \hat v^j_m.
\eeq
Consider the translation-invariant state $|v^j\rangle$ defined by
\beq
|v^j\rangle=\sum_n \int dy \hat{v}^j_n(y) |n y\rangle_y\hsp \hat{v}^j_n(0)=\hat v^j_n.
\eeq
Then the translation generator yields
\bea
P\p|v^j\rangle=\sum_{mn}\int dy \left[-iA_{mn}\hat v_n^{j\prime}(y)+ B_{mn}\hat v^j_n(y)+ C_{mn}y\hat v^j_n(y)
\right]|my\rangle_y=0
\eea
so
\beq
\sum_n\left[-iA_{mn}\hat v_n^{j\prime}(y)+ B_{mn}\hat v^j_n(y)+ C_{mn}y\hat v^j_n(y)
\right]=0.
\eeq
In particular, for small $\epsilon$,
\beq
\hat v^j(\epsilon)=(1-i\epsilon A^{-1}B)\hat v^j. \label{dv}
\eeq

Now let us consider just the component at fixed $y$
\beq
|v^j,y\rangle_y=\sum_n \hat{v}^j_n(y) |n y\rangle_y\hsp |v^j\rangle=\int dy |v^j,y\rangle_y.
\eeq
This is not translation-invariant
\bea
P\p|v^j,0\rangle_y&=&P\p\sum_n \hat{v}^j_n |n 0\rangle_y=\sum_n \hat{v}^j_n P\p\int dy \delta(y) |n y\rangle_y=\sum_n \hat{v}^j_n \lim{\sigma\rightarrow 0} \frac{1}{\sigma\sqrt{2\pi}} P\p\int dy e^{-\frac{y^2}{2\sigma^2}} |n y\rangle_y\nonumber\\
&=&\sum_{mn} \hat{v}^j_n \lim{\sigma\rightarrow 0} \frac{1}{\sigma\sqrt{2\pi}} \int dy e^{-\frac{y^2}{2\sigma^2}}\left[ 
iA_{mn}\frac{y}{\sigma^2}+ B_{mn}+ C_{mn}y
\right] |m y\rangle_y\nonumber\\
&=&\sum_{mn} \hat{v}^j_n \lim{\sigma\rightarrow 0} \frac{1}{\sigma\sqrt{2\pi}} \int dy e^{-\frac{y^2}{2\sigma^2}}\left[ 
iA_{mn}\frac{y}{\sigma^2}+ B_{mn}
\right] |m y\rangle_y.
\eea
In the last equality we used the fact that, as $\sigma\rightarrow 0$, also $y\rightarrow 0$ with $y/\sigma$ fixed.  The coefficient of the $C$ term is $y$, which therefore goes to zero. 

Now let us consider a transformation by a finite distance $\epsilon$.  We will approximate it by the first order transformation inside of the limit, which is legitimate if, when we take $\sigma\rightarrow 0$, we also take $\epsilon/\sigma\rightarrow 0$.   The transformation is
\bea
e^{-i\epsilon P\p}|v^j,0\rangle_y&=&\sum_n \hat{v}^j_n \lim{\sigma\rightarrow 0} \frac{1}{\sigma\sqrt{2\pi}} e^{-i\epsilon P\p}\int dy e^{-\frac{y^2}{2\sigma^2}} |n y\rangle_y\\
&=&\sum_n \hat{v}^j_n \lim{\sigma\rightarrow 0} \frac{1}{\sigma\sqrt{2\pi}} (1-i\epsilon P\p)\int dy e^{-\frac{y^2}{2\sigma^2}} |n y\rangle_y\nonumber\\
&=&\sum_{mn} \hat{v}^j_n \lim{\sigma\rightarrow 0} \frac{1}{\sigma\sqrt{2\pi}} \int dy e^{-\frac{y^2}{2\sigma^2}}\left[\delta_{mn}
+\epsilon A_{mn}\frac{y}{\sigma^2}-i\epsilon B_{mn}
\right] |m y\rangle_y\nonumber\\
&=&\sum_{mn} \hat{v}^j_n \lim{\sigma\rightarrow 0} \frac{1}{\sigma\sqrt{2\pi}} \int dy e^{-\frac{y^2}{2\sigma^2}}\left[\delta_{mn}
+\epsilon \delta_{mn} \lambda_{j}\frac{y}{\sigma^2}-i\epsilon \lambda_j(A^{-1}B)_{mn}
\right] |m y\rangle_y.\nonumber
\eea
Using the expansion (\ref{dv})
\beq
e^{-\frac{(y-\lambda_j\epsilon)^2}{2\sigma^2}}\hat v^j(\lambda_j\epsilon)=e^{-\frac{y^2}{2\sigma^2}}\left[1+\frac{\lambda_j\epsilon y}{\sigma^2} -i\lambda_j\epsilon A^{-1}B
\right]\hat v^j
\eeq
we then conclude
\bea
e^{-i\epsilon P\p}|v^j,0\rangle_y&=&
\sum_{n} \hat v_n^j(\lambda_j\epsilon)  \lim{\sigma\rightarrow 0} \frac{1}{\sigma\sqrt{2\pi}} \int dy e^{-\frac{(y-\lambda_j\epsilon)^2}{2\sigma^2}}  |n y\rangle_y\nonumber\\
&=&\sum_n  \hat v_n^j(\lambda_j\epsilon)|n,\lambda_j\epsilon\rangle_y=|v^j,\lambda_j\epsilon\rangle_y.
\eea
We see that the linearized $y$ coordinates are like the collective $x$ coordinates, except that a translation by $\epsilon$ increases $x$ by $\epsilon$, while it increases $y$ by $\lambda_j\epsilon$.  In particular, this rate depends on the index $j$ on the state being transformed.

\subsection{Linearized Coordinates: Norm}\label{normsec}

To calculate the reduced norm in the linearized $y$ basis, we will need to tie the $y$ and $x$ bases together.  To do this, we will need to match their ordinary normalizations, which we will do by matching their norms.  Both $|v^j\rangle$ and $|v^j,0\rangle$ have infinite norms.  This motivates us to define
\beq
|v^j;\epsilon\rangle_y=\int_0^\epsilon dz e^{-i z P\p}|v^j,0\rangle_y.
\eeq
For small $\epsilon$, its norm is easily calculated
\bea\label{epnorm}
\left||v^j;\epsilon\rangle_y\right|^2
&=&\int_0^\epsilon dz_1\int_0^\epsilon dz_2\ {}_y\langle v^j,0|e^{-i(z_2-z_1) P\p}|v^j,0\rangle_y\\
&=&\sum_{n_1n_2}\int_0^\epsilon dz_1\int_0^\epsilon dz_2  \hat v_{n_1}^{j*}(\lambda_j z_1)\hat v_{n_2}^{j}(\lambda_jz_2){}_y\langle n_1,\lambda_j z_1|n_2,\lambda_jz_2\rangle_y\nonumber\\
&=&\sum_{n_1n_2}\int_0^\epsilon dz_1\int_0^\epsilon dz_2  \hat v_{n_1}^{j*}(\lambda_j z_1)\hat v_{n_2}^{j}(\lambda_jz_2)\delta_{n_1n_2}\delta(\lambda_j (z_1-z_2))
\nonumber\\
&=&\frac{1}{\lambda_j}\sum_{n}\int_0^\epsilon dz \hat v_{n}^{j*}(\lambda_jz)\hat v_{n}^{j}(\lambda_j z).\nonumber
\eea
Now, up to corrections of order $O(\epsilon)$ we can approximate $\hat v(\lambda_jz)=\hat v$.  %Let us normalize the eigenvectors $\hat v^j$ to be orthonormal.  
Then we find
\beq
\left||v^j;\epsilon\rangle_y\right|^2=\frac{1}{\lambda_j}\sum_{n}\hat v_{n}^{j*}\hat v_{n}^{j}\int_0^\epsilon dz =\frac{\epsilon\sum_{n}\hat v_{n}^{j*}\hat v_{n}^{j}}{\lambda_j}=\frac{\epsilon|\hat v^j|^2}{\lambda_j}.
\eeq

\subsection{Collective Coordinates: Norm}
 
 Let us write the same state $|v^j\rangle$ in the collective coordinate basis
\beq
|v^j\rangle=\sum_n v^j_n|e^n\rangle=\sum_n v^j_n\int dx |n x\rangle_x. \label{vdef}
\eeq
Again we can partition this state by the collective coordinate
\beq\label{collcoor}
|v^j,x\rangle_x=\sum_n v^j_n|n x\rangle_x
\eeq
where a translation by $\epsilon$ acts as
\beq
e^{-i\epsilon P\p}|v^j,0\rangle_x=|v^j,\epsilon\rangle_x.
\eeq
To obtain a quantity with a finite norm, we again define
\beq
|v^j;\epsilon\rangle_x=\int_0^\epsilon dx |v^j,x\rangle_x.
\eeq
Calculating as above, its norm is
\beq
\left||v^j;\epsilon\rangle_x\right|^2=\epsilon\sum_n v_n^{j*}v_n^j=\epsilon|v^j|^2.
\eeq

\subsection{Identifying Collective and Linearized Coordinates}

Now we want to identify the $x$ and $y$ bases of the Hilbert space.  Clearly this identification must preserve the norm, and so we choose
\beq
|v^j;\epsilon\rangle_y=\frac{1}{\sqrt{\lambda_j}}\frac{|\hat v^j|}{|v^j|}|v^j;\epsilon\rangle_x.\label{colla}
\eeq
Dividing by $\epsilon$ and taking the limit $\epsilon\rightarrow 0$ this becomes
\beq
\sum_n \hat v^j_n |n 0\rangle_y=|v^j,0\rangle_y=\frac{1}{\sqrt{\lambda_j}}\frac{|\hat v^j|}{|v^j|}|v^j,0\rangle_x=\frac{1}{\sqrt{\lambda_j}}\frac{|\hat v^j|}{|v^j|}\sum_n v_n^j |n 0\rangle_x
\eeq
and so
\beq
\sum_n \frac{\hat v^j_n}{|\hat v^j|} |n 0\rangle_y=\frac{1}{\sqrt{\lambda_j}}\sum_n \frac{v^j_n}{|v^j|} |n 0\rangle_x. 
\eeq
At leading order in $\epsilon$, a translation yields
\beq
\sum_n \frac{\hat v^j_n}{|\hat v^j|} |n, \lambda_j\epsilon\rangle_y=\frac{1}{\sqrt{\lambda_j}}\sum_n \frac{v^j_n}{|v^j|} |n \epsilon\rangle_x.
\eeq

\subsection{Orthogonality}

Recall from Eq.~(\ref{xdef}) that the $|n0\rangle_x$ basis is orthonormal, and from Eq.~(\ref{yort}) that the $|n0\rangle_y$ basis is orthonormal.  As $\hat v^j$ are eigenvectors of a matrix $A$, which we assume to be Hermitian, they will also be orthogonal.  

What about the $v^j$?  Up to a rescaling by $\lambda_j$, these are just $\hat v^j$ written in the $|n0\rangle_x$ basis instead of the $|n0\rangle_y$ basis.  As both bases are orthogonal one expects these to remain orthogonal.  Let us check that this is indeed the case.

Let us take the inner product of Eq.~(\ref{colla}) divided by $\sqrt{\epsilon}$ with itself, at two distinct eigenvalues $j_1\neq j_2$. We denote $\min\{\lambda_{j_1},\lambda_{j_2}\}$ as $\lambda_{\min}$ and $\max\{\lambda_{j_1},\lambda_{j_2}\}$ as $\lambda_{\max}$. The calculation here is similar as in Subsec.~\ref{normsec}. The left hand side yields
%\bea
%\sum_{n_1n_2}{}_y\langle n_1 0|\frac{\hat v^{j_1*}_{n_1}}{|\hat v^{j_1}|}\frac{\hat v^{j_2}_{n_2}}{|\hat v^{j_2}|}|n_2 0\rangle_y=\frac{\sum_n \hat v^{j_1*}_n\hat v^{j_2}_n}{|\hat v^{j_1}||\hat v^{j_2}|}=0
%\eea
\bea
\frac{1}{\epsilon}{}_y\langle v^{j_1};\epsilon|v^{j_2};\epsilon\rangle_y
&=&\frac{1}{\epsilon}\int_0^\epsilon dz_1\int_0^\epsilon dz_2\ {}_y\langle v^{j_1},0|e^{-i(z_2-z_1) P\p}|v^{j_2},0\rangle_y\\
&=&\frac{1}{\epsilon}\sum_{n_1n_2}\int_0^\epsilon dz_1\int_0^\epsilon dz_2  \hat v_{n_1}^{j_1*}(\lambda_j z_1)\hat v_{n_2}^{j_2}(\lambda_jz_2){}_y\langle n_1,\lambda_{j_1} z_1|n_2,\lambda_{j_2} z_2\rangle_y\nonumber\\
&=&\frac{1}{\epsilon}\sum_{n_1n_2}\int_0^\epsilon dz_1\int_0^\epsilon dz_2  \hat v_{n_1}^{j_1*}(\lambda_j z_1)\hat v_{n_2}^{j_2}(\lambda_jz_2)\delta_{n_1n_2}\delta(\lambda_{j_1}z_1 - \lambda_{j_2}z_2)
\nonumber\\
&=&\frac{1}{\epsilon\lambda_{j_1}\lambda_{j_2}}\sum_{n_1n_2}\int_0^\epsilon d(\lambda_{j_1}z_1) \int_0^\epsilon d(\lambda_{j_2}z_2) \hat v_{n_1}^{j_1*}(\lambda_{j_1}z_1)\hat v_{n_2}^{j_2}(\lambda_{j_2} z_2)\delta_{n_1n_2}\delta(\lambda_{j_1}z_1 - \lambda_{j_2}z_2)\nonumber\\
&=&\frac{1}{\epsilon\lambda_{j_1}\lambda_{j_2}}\sum_{n}\int_0^{\lambda _{j_1}\epsilon} d \tilde{z}_1 \int_0^{\lambda_{j_2}\epsilon} d\tilde{z}_2 \hat v_{n}^{j_1*}(\tilde{z}_1)\hat v_{n}^{j_2}(\tilde{z}_2)\delta(\tilde{z}_1-\tilde{z}_2)\nonumber\\
&=&\frac{1}{\epsilon\lambda_{\min}\lambda_{\max}}\sum_{n}\int_0^{\lambda _{\min}\epsilon} d \tilde{z}  \hat v_{n}^{j_1*}(\tilde{z})\hat v_{n}^{j_2}(\tilde{z}).\nonumber
\eea
Again as in Subsec.~\ref{normsec}, up to corrections of order $O(\epsilon)$ we can approximate $\hat v(\tilde{z})=\hat v$. Then we find
\beq
\frac{1}{\epsilon}{}_y\langle v^{j_1};\epsilon|v^{j_2};\epsilon\rangle_y=\frac{1}{\epsilon\lambda_{\min}\lambda_{\max}}\sum_{n}\hat v_{n}^{j_1*}\hat v_{n}^{j_2}\int_0^{\lambda _{\min}\epsilon} d \tilde{z}=\frac{\sum_{n}\hat v_{n}^{j_1*}\hat v_{n}^{j_2}}{\lambda_{\max}}=0
\eeq
where the last equality used the orthogonality of the $\hat v$. 

 Here we assumed that all $\lambda_j>0$.  In the case of interest of quantum kinks, $A$ will be a positive scalar plus a correction suppressed by a power of the coupling, so this is the case at small coupling.

The calculation of the right hand side is similar
\bea
0&=&\frac{1}{\epsilon\sqrt{\lambda_{j_1} \lambda_{j_2}}}\frac{|\hat v^{j_1}||\hat v^{j_2}|}{|v^{j_1}||v^{j_2}|}{}_x\langle v^{j_1};\epsilon|v^{j_2};\epsilon\rangle_x\\
&=&\frac{1}{\epsilon\lambda_{j_1} \lambda_{j_2}}\int_0^\epsilon dx_1\int_0^\epsilon dx_2\ {}_x\langle v^{j_1},x_1|v^{j_2},x_2\rangle_x\nonumber\\
&=&\frac{1}{\epsilon\lambda_{j_1} \lambda_{j_2}} \sum_{n_1n_2}v_{n_1}^{j_1*} v_{n_2}^{j_2}\int_0^\epsilon dx_1\int_0^\epsilon dx_2\ {}_x\langle n_1,x_1|n_2,x_2\rangle_x\nonumber\\
&=&\frac{1}{\epsilon\lambda_{j_1} \lambda_{j_2}} \sum_{n_1n_2}v_{n_1}^{j_1*} v_{n_2}^{j_2}\int_0^\epsilon dx_1\int_0^\epsilon dx_2\ \delta_{n_1n_2}\delta(x_1-x_2)\nonumber\\
&=&\frac{1}{\epsilon\lambda_{j_1} \lambda_{j_2}} \sum_{n}v_{n}^{j_1*} v_{n}^{j_2}\int_0^\epsilon dx=\frac{ \sum_{n}v_{n}^{j_1*} v_{n}^{j_2}}{\lambda_{j_1} \lambda_{j_2}} \nonumber
\eea
where from the 2nd line to the 3rd line we used Eq.~(\ref{collcoor}).  In passing from the first line to the second, we used the identity~(\ref{vhatv}) which will be proved momentarily. So the $v$ are also orthogonal
\beq
\sum_n v^{j_1*}_n v^{j_2}_n=0.
\eeq

%where the first equality used Eq.~(\ref{yort}) and the second used the orthogonality of the $\hat v$.  Thus the right hand side also vanishes
%\bea
%0=\frac{1}{\sqrt{\lambda_{j_1}\lambda_{j_2}}}\sum_{n_1n_2}{}_x\langle n_1 0|\frac{ v^{j_1*}_{n_1}}{|v^{j_1}|}\frac{ v^{j_2}_{n_2}}{| v^{j_2}|}|n_2 0\rangle_x=\frac{1}{\sqrt{\lambda_{j_1}\lambda_{j_2}}}\frac{\sum_n v^{j_1*}_n v^{j_2}_n}{|v^{j_1}||v^{j_2}|}.
%\eea
%The eigenvalues and norms are finite and so the $v$ are also orthogonal
%\beq
%\sum_n v^{j_1*}_n v^{j_2}_n=0.
%\eeq

\subsection{Linearized Coordinates: Reduced Norm}

Finally we are ready to compute the reduced norm of $|v^j\rangle$.  The reduced norm squared is defined to be $|v^j|^2$.  The following manipulations are valid at small $y$, where the $y$-coordinate perturbation theory is valid
\bea
|v^j\rangle&=&\sum_n\int dy \hat v^j_n(y)|n y\rangle_y=\frac{1}{\sqrt{\lambda_j}}|\hat v^j|\sum_n \frac{v^j_n}{|v^j|}\int dy|n, y/\lambda_j\rangle_x\\
&=&{\sqrt{\lambda_j}}|\hat v^j|\sum_n \frac{v^j_n}{|v^j|}\int dx|n, x\rangle_x={\sqrt{\lambda_j}}|\hat v^j|\sum_n \frac{v^j_n}{|v^j|}|e^n\rangle.
\nonumber
\eea
Matching to Eq.~(\ref{vdef}), we find
\beq\label{vhatv}
\sqrt{\lambda_j}|\hat v^j|=|v^j|.
\eeq

The reduced norm is therefore
\bea\label{rednorm}
{}_{\rm{}}\langle v^j|v^j\rangle_{\rm{red}}&=&\lambda_j |\hat v^j|^2 \sum_{n_1n_2}\frac{v^{j*}_{n_1}v^{j}_{n_2}}{|v^j|^2}{}_{\rm{}}\langle e^{n_1}|e^{n_2}\rangle_{\rm{red}}\\
&=&\lambda_j |\hat v^j|^2 \sum_{n_1n_2}\frac{v^{j*}_{n_1}v^{j}_{n_2}}{|v^j|^2}\delta_{n_1n_2}=\lambda_j|\hat{v}^j|^2=\sum_{mn}\hat v^{j*}_m A_{mn}\hat v^j_n.
\nonumber
\eea
Similarly, if $j\neq k$ then the reduced inner product is 
\bea
{}_{\rm{}}\langle v^j|v^k\rangle_{\rm{red}}&=&\sqrt{\lambda_j\lambda_k}|\hat v^j||\hat v^k| \sum_{n_1n_2}\frac{v^{j*}_{n_1}v^{k}_{n_2}}{|v^j||v^k|}\delta_{n_1n_2}\\
&=&\sqrt{\lambda_j\lambda_k}|\hat v^j||\hat v^k| \sum_{n}\frac{v^{j*}_{n}v^{k}_{n}}{|v^j||v^k|}=0=\lambda_k \sum_{n}\hat v^{j*}_n\hat v^k_n=\sum_{mn}\hat v^{j*}_m A_{mn}\hat v^k_n.\nonumber
\eea

Now consider any two translation-invariant states $|\phi\rangle$ and $|\psi\rangle$.  Assume that $A$ is Hermitian, so that its eigenvectors are a basis of the vector space generated by the $|e^n\rangle$.  Then
\bea
|\psi\rangle&=&\sum_n \int dy\hat\psi_n(y)|ny\rangle_y=\sum_n \int dy\left[\hat\psi_n+O(y)\right]|ny\rangle_y=\sum_n \hat\psi_n|n\rangle_y=\sum_{jn}\hat\psi_n\left(\hat v^{-1}\right)^j_n |v^j\rangle\nonumber\\
|\phi\rangle&=&\sum_{jn}\hat\phi_n\left(\hat v^{-1}\right)^j_n |v^j\rangle\label{2trans}
\eea
where we have defined
\beq
|n\rangle_y=\int dy |ny\rangle_y. \label{ndef}
\eeq
Here we have dropped the term of order $O(y)$, as translation-invariance implies that the matching of the $y$ and $x$ kets can be applied at any value of $y$, and we apply it at $y=0$ where the $O(y)$ correction vanishes.

Their reduced inner product is
\bea
\langle \phi|\psi\rangle_{\rm{red}}&=&\sum_{n_1n_2j_1j_2}\hat\phi_{n_1}^*\left(\hat v^{*-1}\right)^{j_1}_{n_1}\hat\psi_{n_2}\left(\hat v^{-1}\right)^{j_2}_{n_2}
\langle v^{j_1}|v^{j_2}\rangle_{\rm{red}}\label{qmpadr}\\
&=&\hat\phi^* \left(\hat v^*\right)^{-1}\hat v^* A \hat v \hat v^{-1}\hat \psi=\hat\phi^* A \hat\psi.\nonumber
\eea

\subsection{Interpretation}

Let us pause to interpret our result (\ref{qmpadr}).  The inner product of
\beq
|\psi\rangle=\sum_n \int dy \hat\psi_n(y)|ny\rangle_y\hsp
|\phi\rangle=\sum_n \int dy \hat\phi_n(y)|ny\rangle_y.
\eeq
is infrared divergent, due to the $y$ integral. However these inner products only appear in ratios, so it is sufficient to consider the inner product per unit of translation, dividing through by the volume of the translation group. 

Translation symmetry acts transitively on the $y$ coordinate, leaving the states invariant.  Therefore we can calculate this inner product density in a neighborhood of any fixed $y$.  Consider $y=0$.  Close to this point, we can approximate
\beq
|\psi\rangle=\sum_n \hat\psi_n \int dy |ny\rangle_y\hsp
|\phi\rangle=\sum_n \hat\phi_n \int dy |ny\rangle_y.
\eeq

The inner product is still divergent, but close to the basepoint $y=0$ it factorizes
\beq
{}_y\langle \phi y_1|A|\psi y_2\rangle_y=\langle \phi|\psi\rangle_{\rm{red}}\delta(y_1-y_2). \label{amp}
\eeq
We learn that the reduced inner product $\langle \phi|\psi\rangle_{\rm{red}}$ is given by the vector inner product of $\hat\psi$ and $\hat\phi$ with a Jacobian factor, $A$, resulting from the difference between the translation operator $P\p$ and a rigid shift in $y$.  Thus to evaluate the reduced inner product, which intuitively contains an invariant $\delta(x_1-x_2)$ in the collective coordinate, one first decomposes $\delta(x_1-x_2)=A\delta(y_1-y_2)$ and then amputates the $\delta(y_1-y_2)$ which was added to decompose the left hand side.

%The summed expression has a reduced norm of $1$, and so the reduced norm of $|v^j\rangle$ is $\sqrt{\lambda_j}|\hat v^j|$.

%If $A$ is Hermitian then these eigenvectors generate the entire space and so we may invert the eigenvector matrix.

%Let us define the matrix $N(y)$ by
%\beq
%N(y)=A+T(y)
%\eeq
%where, for all $P\p$-invariant states,
%\beq
%T(y)\hat\psi\p(y)=(B+yC)\hat\psi(y)=-A\hat \psi\p(y)
%\eeq
%so that
%\beq
%N(y)\hat\psi\p(y)=0.
%\eeq
%Then our invariant reduced norm is
%\beq
%\langle \phi|\psi\rangle_{\rm{red}}=\sum_{mn}\hat\phi_m^*(y)N_{mn}\hat\psi_n(y).
%\eeq

%But what is $T(y)$?

%Now let us define $\hat x$ to be any operator that transforms as
%\beq
%[P\p,\hat x]=-i.
%\eeq
%If $x_0$ commutes with $P\p$, then we may construct
%\beq
%y(\hat{x})=
%\eeq

\section{Reduced Norms for Quantum Kinks} \label{qftsez}

\subsection{Notation}

To pass from quantum mechanics to the case of a quantum field theory admitting quantum kinks, we make the following replacements.  First, the discrete quantum number $n$ is replaced by symmetrized $n$-tuples of continuum and shape normal mode labels $k$.  Here $k$ is a real number for continuum modes, and a discrete index for shape modes.  Now $n\geq 0$ and these states represent the $n$-meson Fock space in the kink sector.

We let $\phi_0$ be the operator whose eigenvalue is $y$.  Its dual momentum we recall is $\pi_0$.  We introduce the shorthand
\beq
\Delta_{ij}=\int dx \g_i(x) \g\p_j(x)
\eeq
where $i$ and $j$ run over the zero mode $B$, as well as continuum modes $k$ and shape modes $S$.

The translation operator $P\p$ is now given by
\bea
P\p&=&P+\sqrt{Q_0}\pi_0 \label{ppk}\\
P&=&\ppin{k}\Delta_{kB}\left[i\phi_0\left(-\ok{}B^\ddag_k+\frac{B_{-k}}{2}\right)+\pi_0\left(B^\ddag_k+\frac{B_{-k}}{2\ok{}}\right)\right]\nonumber\\
&&+i\ppink{2}\Delta_{k_1k_2}\left[\frac{\ok{2}-\ok{1}}{2}B^\ddag_{k_1}B^\ddag_{k_2}-\frac{1}{2}\left(1+\frac{\ok{1}}{\ok{2}}\right)B^\ddag_{k_1}B_{-k_2}+\frac{\ok{1}-\ok{2}}{8\ok{1}\ok{2}}B_{-k_1}B_{-k_2}
\right].\nonumber
\eea
Intuitively $P$ is the momentum operator for the mesons while $\sqrt{Q_0}\pi_0$ is the momentum operator for the kink.  Only $P\p$ is conserved.  Recalling our old decomposition (\ref{pp})
\beq
P\p=A\pi_0+B+C\phi_0
\eeq
we can match the $\pi_0$ coefficient in (\ref{ppk}) to obtain
\beq
A=\sqrt{Q_0}+ \ppin{k}\Delta_{kB}\left( B^\ddag_k+\frac{B_{-k}}{2\ok{}}\right).
\eeq

We decompose states as
\bea
|\psi\rangle&=&\sum_{m,n=0}^\infty |\psi\rangle^{mn}\hsp
|k_1\cdots k_n\rangle_0=\Bd1\cdots\Bd n\vac_0
\nonumber\\
|\psi\rangle^{mn}&=&\phi_0^m\ppink{n}\gamma_\psi^{mn}(k_1\cdots k_n)|k_1\cdots k_n\rangle_0\hsp \vac_0=\int dy |y\rangle_y.
 \label{gameqa}
\eea

To make contact with the decomposition in Sec.~\ref{qmsez}, note that
\bea\label{yk0}
|\psi\rangle^{mn}&=&\int dy |y,\psi\rangle_y^{mn}\hsp
|y,\psi\rangle^{mn}_y
=y^m\ppink{n}\gamma_\psi^{mn}(k_1\cdots k_n)|y,k_1\cdots k_n\rangle_y\nonumber\\
|y,k_1\cdots k_n\rangle_y&=&\Bd1\cdots\Bd n|y\rangle_y.
\eea
We see that here the role which was played by $\hat{\psi}_n(y)$ in quantum mechanics is now played by
\beq
\hat{\psi}_n(y) \rightarrow \sum_m \gamma_\psi^{mn}(k_1\cdots k_n) y^m.
\eeq
The discrete $n$ quantum number is replaced by an $n$-tuple of shape and continuum mode indices $k$
\beq
|n y\rangle_y \rightarrow |y,k_1\cdots k_n\rangle_y\hsp
\sum_n \rightarrow \sum_n \ppink{n}. \label{nymap}
\eeq

Recall that, in quantum mechanics, the coefficient at the origin was $\hat\psi_n=\hat\psi_n(0)$.  Similarly, setting $y=0$ here we obtain $\gamma^{0n}$
\beq
\hat{\psi}_n \rightarrow  \gamma_\psi^{0n}(k_1\cdots k_n) .
\eeq

\subsection{The Reduced Inner Product}

Let us first evaluate
\bea
A|0,\psi\rangle_y^{0n}&=&\left(
\sqrt{Q_0}+ \ppin{k}\Delta_{kB}\left( B^\ddag_k+\frac{B_{-k}}{2\ok{}}\right)
\right)|0,\psi\rangle_y^{0n}\\
&=&\ppink{n}\gamma_\psi^{0n}(k_1\cdots k_n)
\left(
\sqrt{Q_0}+ \ppin{k\p}\Delta_{k\p B}\left( B^\ddag_{k\p}+\frac{B_{-k\p}}{2\okp{}}\right)\right)
|0,k_1\cdots k_n\rangle_y
\nonumber\\
&=&\sqrt{Q_0}|0,\psi\rangle_y^{0n}+\ppink{n+1}\gamma_\psi^{0n}(k_1\cdots k_n)\Delta_{k_{n+1},B}|0,k_1\cdots k_{n+1}\rangle_y\nonumber\\
&&+n\ppink{n}\gamma_\psi^{0n}(k_1\cdots k_n)\frac{\Delta_{-k_n,B}}{2\ok n}|0,k_1\cdots k_{n-1}\rangle_y
\nonumber
\eea
where, in the last line, we have assumed that $\gamma_\psi^{0n}$ is symmetrized over its arguments $k_i$.  Summing over $n$ one arrives at
\bea\label{A0psi}
A\sum_n|0,\psi\rangle_y^{0n}&=&\sum_n \ppink{n}
\left[ \sqrt{Q_0}\gamma_\psi^{0n}(k_1\cdots k_n)+
\gamma_\psi^{0,n-1}(k_1\cdots k_{n-1})\Delta_{k_n,B}
\right.\nonumber\\
&&\left.
+(n+1)\ppin{k_{n+1}}\gamma_\psi^{0,n+1}(k_1\cdots k_{n+1})\frac{\Delta_{-k_{n+1}, B}}{2\ok{n+1}}
\right]
|0,k_1\cdots k_{n}\rangle_y.
\eea

Using the oscillator algebra satisfied by $B$ and $B^\ddag$ and ${}_y\langle y_1|y_2\rangle_y=\delta(y_1-y_2)$ one finds the inner products of
\beq
|a_i,y_i\rangle=\ppink{n}a_i(k_1\cdots k_n)|y_i,k_1\cdots k_n\rangle_y
\eeq
where $a_i$ is symmetric in its arguments, to be
\beq
\langle a_1,y_1|a_2,y_2\rangle=n!\delta(y_1-y_2)\ppink{n}\frac{a_1^*(k_1\cdots k_n)a_2(k_1\cdots k_n)}{\prod_{i=1}^n(2\ok{i})}. \label{qfti}
\eeq

Finally we want to generalize the reduced inner product (\ref{qmpadr}) to quantum field theory.  To do this, we need to generalize the vector inner product of the $\hat v$ vectors.  Our definition is that this inner product is to be interpreted as the full inner product (\ref{qfti}) in quantum field theory, without the $\delta(y_1-y_2)$.  This is just the quantum field theory generalization of Eq.~(\ref{amp}).%Indeed, the same prescription would apply in quantum mechanics, where the reduced inner product of two $\hat v$ vectors is equal to ${}_y\langle v^1 y_1|A|v^2 y_2\rangle_y$ with the $\delta(y_1-y_2)$ amputated as a result of Eq.~(\ref{yort}).

We then obtain our master formula for the reduced inner product in quantum field theory\bea
{}_{\rm{}}{}\langle \phi|\psi\rangle_{\rm{red}}&=&\sum_{n_1n_2}{}_{\ \ y}^{0n_1}\langle y_1,\phi|A|y_2,\psi\rangle_y^{0n_2}|_{{\rm Coefficient\ of}\ \delta(y_1-y_2){\rm \ at}\ y_1=0} \label{padqft}
\\
%&&\hspace{-2cm}=\sum_n {}_{\ \ y}^{0n_1}\langle 0,\phi|\left[ 
%\sqrt{Q_0}|0,\psi\rangle_y^{0n}
%+ \ppin{k}\Delta_{kB} B^\ddag_k|0,\psi\rangle_y^{0,n-1}
%+\ppin{k}\Delta_{kB}\frac{B_{-k}}{2\ok{}}|0,\psi\rangle_y^{0,n+1}
%\right]
%\nonumber\\
&=&%\hspace{-2cm}
\sum_n n! \ppink{n}\frac{\gamma_\phi^{0n*}(k_1\cdots k_n)}{\prod_{i=1}^n(2\ok{i})}
\left[ \sqrt{Q_0}\gamma_\psi^{0n}(k_1\cdots k_n)+
\gamma_\psi^{0,n-1}(k_1\cdots k_{n-1})\Delta_{k_n,B}
\right.\nonumber\\
&&\left.
+(n+1)\ppin{k_{n+1}}\gamma_\psi^{0,n+1}(k_1\cdots k_{n+1})\frac{\Delta_{-k_{n+1},B}}{2\ok{n+1}}
\right].
\nonumber
\eea
This is our main result.  We remind the reader that all $\gamma$ must be symmetrized in their arguments before this formula applied, or else the $n!$ and $(n+1)!$ factors should be replaced with sums over $S_n$ and $S_{n+1}$ permutations.  The second and third terms in the square brackets are the Jacobian terms resulting from the off-diagonal part of $A$.   Both are proportional to $\Delta_{kB}$, which describes the mixing between the zero mode and the normal modes as the kink moves.

In the next section we will see that this formula satisfies some basic consistency checks.  For example, we will see that the reduced inner product of a 0-meson and 1-meson state vanishes, whereas one would obtain a nonzero result if one did not include the Jacobian terms in (\ref{padqft}). 

\section{Examples of Reduced Norms} \label{exsez}

In this section we will calculate the reduced norms of the kink ground state and also a kink with one excitation, which can be a continuum meson or a bound shape mode.  We will show that, up to corrections which are suppressed, with respect to the leading term, by a quantity of order $O(\lambda)$
\beq
\langle 0\vac_{\rm{red}}=\sqrt{Q_0}+O(\sl)\hsp
\langle\kt_1|\kt_2\rangle_{\rm{red}}=\frac{2\pi\delta(\kt_1-\kt_2)}{2\okt 1}\sqrt{Q_0}+O(\sl). \label{grred}
\eeq
Had there been corrections of order $O(\lambda^0)$, which would be suppressed with respect to the leading term by only one power of $\sl$, this would have invalidated calculations in Refs.~\cite{alberto} and \cite{memult}.

We will decompose each $\gamma^{mn}$ as
\beq
\gamma^{mn}=\sum_i Q_0^{-i/2}\gamma_i^{mn}.
\eeq

\subsection{The Reduced Norm of the Ground State}

The kink ground state, at subleading order, is characterized by the coefficients\footnote{These coefficients were calculated in Ref.~\cite{me2loop}.  As a result of a sign difference in the convention (\ref{gb}) for $\g_B(x)$, here the meson and kink momentum contributions to $P\p$ have a different relative sign.  This changes the signs of all $\Delta$ terms in all coefficients. Also, the convention for $\v3$ here differs by a factor of $\sqrt{\lambda}$.}
\bea
\gamma_0^{00}&=&1\hsp
\gamma_1^{12}(k_1,k_2)=\frac{\left(\ok 2-\ok 1\right)\Delta_{k_1k_2}}{2}\hsp
\gamma_1^{21}(k_1)=-\frac{\omega_{k_1}\Delta_{k_1B}}{2}\\
\gamma_1^{01}(k_1)&=&-\frac{\Delta_{k_1B}}{2}-\frac{\sqrt{\lambda Q_0}}{2}\frac{V_{\I k_1}}{\ok{1}}\hsp
\gamma_1^{03}(k_1,k_2,k_3)=-\frac{\sqrt{\lambda Q_0}}{6}\frac{V_{k_1k_2k_3}}{\ok{1}+\ok{2}+\ok{3}}\nonumber
\eea
where we have defined
\bea
V_{k_1\cdots k_n}&=&\int dx \V{n} \g_{k_1}(x)\cdots  \g_{k_n}(x)\\
V_{\I k_1\cdots k_n}&=&\int dx \V{n+2} \I(x) \g_{k_1}(x)\cdots\g_{k_n}(x)\nonumber\\
\I(x)&=&\pin{k}\frac{\left|{\g}_{k}(x)\right|^2-1}{2\omega_k}+\sum_S \frac{\left|{\g}_{S}(x)\right|^2}{2\omega_k}.
\nonumber
\eea
In the first two lines, the $k_i$ run over not just the continous momenta, but also the shape modes.

%Let us calculate its reduced norm.  First, note that each term has a different number of mesons except for $\gamma_1^{21}$ and $\gamma_1^{01}$, which each have one meson.  Thus, except for that pair, the corresponding terms are all orthogonal.  Fixing the translation invariance by setting $y=0$, we eliminate all terms with $m\neq 0$.

%Also, we have seen in Eq.~(\ref{padr}) that for all terms except for the terms that differ by a single meson, the determinant factor in our definition (\ref{redk}) of the reduced norm is just $y\p(Y)=\sqrt{Q_0}$.  The only two terms that differ by a single meson are $\gamma_0^{00}$ and $\gamma_1^{01}.$

%We therefore find the reduced norms
The reduced norm can be written as a sum of three terms corresponding to $n=0$,\ $1$\ and $3$\ in Eq.~(\ref{padqft})
\bea
|\vac|^2_{\rm{n,red}}&=&n! \ppink{n}\frac{\gamma^{0n*}(k_1\cdots k_n)}{\prod_{i=1}^n(2\ok{i})}
\left[ \sqrt{Q_0}\gamma^{0n}(k_1\cdots k_n)+
\gamma^{0,n-1}(k_1\cdots k_{n-1})\Delta_{k_n,B}
\right.\nonumber\\
&&\left.
+(n+1)\ppin{k_{n+1}}\gamma^{0,n+1}(k_1\cdots k_{n+1})\frac{\Delta_{-k_{n+1}, B}}{2\ok{n+1}}
\right].
\eea

These summands are
\bea
|\vac|^2_{\rm{0,red}}&=&
 \sqrt{Q_0}+\ppin{k_1}\gamma^{01}(k_1)\frac{\Delta_{-k_1 B}}{2\ok{1}}\label{0rednorm}\\
 &=&
 \sqrt{Q_0}-\frac{1}{4\sqrt{Q_0}}\ppin{k_1}\left[ 
 {\Delta_{k_1 B}}{}+{\sqrt{\lambda Q_0}}{}\frac{V_{\I k_1}}{\ok{1}}
 \right]\frac{\Delta_{-k_1 B}}{\ok{1}}\nonumber\\
|\vac|^2_{\rm{1,red}}&=& \ppin{k_1}\frac{\gamma^{01*}(k_1)}{2\ok{1}}
\left[ \sqrt{Q_0}\gamma^{01}(k_1)+
\Delta_{k_1B}
\right] \nonumber\\
&=& \frac{1}{8\sqrt{Q_0}}\ppin{k_1}\left[\frac{\lambda Q_0 |V_{\I k_1}|^2}{\ok{1}^3}-\frac{|\Delta_{k_1B}|^2}{\ok{1}}\right]\nonumber\\
|\vac|^2_{\rm{3,red}}&=&\frac{3\sqrt{Q_0}}{4} \ppink{3}\frac{\left|\gamma^{03}(k_1,k_2, k_3)\right|^2}{\prod_{i=1}^3\ok{i}}\nonumber\\
&=&\frac{\lambda\sqrt{Q_0}}{48} \ppink{3}\frac{\left|V_{k_1k_2 k_3}\right|^2}{\ok  1\ok 2\ok 3(\ok{1}+\ok 2+\ok 3)^2}.\nonumber
\eea

Altogether we find
%\bea
%|\vac|^2_{\rm{red}}&=&\sqrt{Q_0}-\frac{1}{8\sqrt{Q_0}}\ppin{k_1}\frac{1}{\ok 1}\left|{\Delta_{k_1 B}}{}+{\sqrt{\lambda Q_0}}{}\frac{V_{\I k_1}}{\ok{1}}\right|^2\\
%&&+\frac{\lambda\sqrt{Q_0}}{48} \ppink{3}\frac{\left|V_{k_1k_2 k_3}\right|^2}{\ok  1\ok 2\ok 3(\ok{1}+\ok 2+\ok 3)^2}.\nonumber
%\eea
\bea
|\vac|^2_{\rm{red}}&=&\sqrt{Q_0}+\frac{1}{8\sqrt{Q_0}}\ppin{k_1}\frac{1}{\ok 1}\left({\sqrt{\lambda Q_0}}{}\frac{V_{\I k_1}}{\ok{1}}+{\Delta_{k_1 B}}{}\right)\left({{\sqrt{\lambda Q_0}}{}\frac{V_{\I -k_1}}{\ok{1}}{}-3\Delta_{-k_1 B}}\right)\nonumber\\
&&+\frac{\lambda\sqrt{Q_0}}{48} \ppink{3}\frac{\left|V_{k_1k_2 k_3}\right|^2}{\ok  1\ok 2\ok 3(\ok{1}+\ok 2+\ok 3)^2}.
\eea
Note that we have adapted the convention $\gamma_2^{00}=0$.  Another convention for $\gamma_2^{00}$ would have resulted in a different norm.  This is the lowest order manifestation of the freedom in choosing the overall normalization, which is already present quantum mechanics.  Clearly there is such a freedom for every state.  Although the norms of all states are a matter of convention, the determination of the norm for a given convention is physically relevant, as the convention then fixes all of the reduced inner products.

%Summing these, one arrives at the norm of the approximation $\vac_0+\vac_1$ to the kink ground state $\vac$.

\subsection{Inner Product of Zero and One-Meson State}

\subsubsection{The One-Meson States up to $O(\sqrt{\lambda})$}

Now let us consider a one-meson Hamiltonian eigenstate $|\kt\rangle$.  At leading order, it is $|\kt\rangle_0$, characterized by
\beq
\gamma_{0\kt}^{01}(k_1)=2\pi\delta(k_1-\kt). \label{g0}
\eeq
The next order corrections\footnote{They were calculated in Ref.~\cite{menormal}, again with the sign flip for all $\Delta$ symbols and $\sqrt{\lambda}$ for $V$ symbols resulting from the convention (\ref{gb}).} are summarized by the corresponding symbols $\gamma_{1\kt}^{mn}$
\bea
\gamma_{1\kt}^{11}(k_1)&=&\frac{1}{2}\Delta_{-\kt k_1}\left(1+\frac{\ok1}{\omega_{\kt}}\right)\hsp
\gamma_{1\kt}^{13}(k_1,k_2,k_3)=\ok3\Delta_{k_2k_3}2\pi\delta(k_1-\kt)\label{gammakt}\\
\gamma_{1\kt}^{22}(k_1,k_2)&=&-\frac{\ok2}{2}\Delta_{k_2 B}2\pi\delta(k_1-\kt)\hsp
\gamma_{1\kt}^{00}= \frac{\sqrt{Q_0\lambda}V_{\I, -\kt}}{4\omega^2_{\kt}}-\frac{\Delta_{-\kt B}}{4\omega_\kt}\nonumber\\
\gamma_{1\kt}^{02}(k_1,k_2)&=& -\frac{2\pi\delta(k_2-\kt)}{4}\left(\Delta_{k_1 B}+\sqrt{Q_0\lambda}\frac{V_{\I  k_1}}{\ok1}\right)+\frac{\sqrt{Q_0\lambda}V_{-\kt k_1 k_2}}{4\omega_\kt\left(\omega_\kt-\ok1-\ok2\right)}\nonumber\\
&&-\frac{2\pi\delta(k_1-\kt)}{4}\left(\Delta_{k_2 B}+\sqrt{Q_0\lambda}\frac{V_{\I  k_2}}{\ok 2}\right)\nonumber\\
\gamma_{1\kt}^{04}(k_1\cdots k_4)&=& -\frac{\sqrt{Q_0\lambda}V_{k_1 k_2 k_3}}{6\sum_{j=1}^3 \ok{j}}2\pi\delta(k_4-\kt)\hsp
\gamma_{1\kt}^{20}=\frac{1}{4}\Delta_{-\kt B}.\nonumber
\eea

\subsubsection{The Inner Product}

Let us calculate the reduced inner product of the kink ground state $\vac$ and a one-kink one-meson state $|\kt\rangle$ up to order $O(\lambda^0)$.  There are two contributions
\bea
\langle 0|\kt\rangle_{\rm{n,red}}&=&n! \ppink{n}\frac{\gamma^{0n*}(k_1\cdots k_n)}{\prod_{i=1}^n(2\ok{i})}
\left[ \sqrt{Q_0}\gamma^{0n}_{\kt}(k_1\cdots k_n)+
\gamma^{0,n-1}_{\kt}(k_1\cdots k_{n-1})\Delta_{k_n,B}
\right.\nonumber\\
&&\left.
+(n+1)\ppin{k\p}\gamma_{\kt}^{0,n+1}(k_1\cdots k_n,k\p)\frac{\Delta_{-k\p B}}{2\okp{}}
\right]
\eea
at this order.  These are
\bea
\langle 0|\kt\rangle_{\rm{0,red}}&=& \sqrt{Q_0}\gamma^{00}_{\kt}+\ppin{k\p}\gamma_{\kt}^{01}(k\p)\frac{\Delta_{-k\p B}}{2\okp{}}=\frac{\sqrt{Q_0\lambda}V_{\I -\kt}}{4\omega^2_{\kt}}+\frac{\Delta_{-\kt B}}{4\omega_\kt}\\
\langle 0|\kt\rangle_{\rm{1,red}}&=& \ppin{k_1}\frac{\gamma^{01*}(k_1)}{2\ok{1}}
\sqrt{Q_0}\gamma^{01}_{\kt}(k_1)=\frac{\gamma_1^{01*}(\kt)}{2\okt{}}=
-\frac{\Delta_{-\kt B}}{4\okt{}}-\frac{\sqrt{\lambda Q_0}}{2}\frac{V_{\I -\kt}}{2\okt{}^2}.\nonumber
\eea
These cancel precisely, leaving
\beq
\langle 0|\kt\rangle_{\rm{red}}=0
\eeq
at order $O(\lambda^0)$.  This is to be expected, as $\vac$ and $|\kt\rangle$ are eigenstates of  $H\p$ with distinct eigenvalues.  Note that the off-diagonal terms in $A$ contributed to $\langle 0|\kt\rangle_{\rm{0,red}}$ and were necessary for the reduced inner product to respect this orthogonality.

\subsection{Inner Product of Two One-Meson States}

We next turn our attention to the reduced inner product of two one-meson, one-kink states, $|\kt_1\rangle$ and $|\kt_2\rangle$, at $O(\sl)$.

\subsubsection{A Coefficient at $O(\lambda)$}

In addition to the $O(\lambda^0)$ coefficient given in Eq.~(\ref{g0}) and the $O(\sl)$ coefficients given in Eq.~(\ref{gammakt}), we will also need the $O(\lambda)$ coefficient $\gamma_{2\kt}^{01}(k_1)$ at $k_1\neq\kt$.  To calculate this, we use the eigenvalue equation
\beq
(H\p-E)|\kt\rangle=0.
\eeq
At order $O(\lambda)$ this consists of five terms
\beq
0=H\p_4|\kt\rangle_0+H\p_3|\kt\rangle_1+H\p_2|\kt\rangle_2-E_2|\kt\rangle_0-E_1|\kt\rangle_2. \label{schrod}
\eeq
Here $E_n$ is the $O(\lambda^{n-1})$ term in the energy of the 1-kink, 1-meson state $|\kt\rangle$.  In particular
\beq
E_1=\okt{}\hsp 
E_2=\sigma_\kt\hsp
H\p_2=\frac{\pi_0^2}{2}+\ppin{k}\ok{}\Bd{}B_k
\eeq
where $\sigma_\kt$ was calculated in Ref.~\cite{menormal}.  Note that, since $H\p_0=E_0=Q_0$ is a scalar, the $H_0$ and $E_0$ terms that one may be tempted to include would cancel.

We will impose Eq.~(\ref{schrod}) on the terms which are independent of $\phi_0$ and contain one meson, in other words the $m=0$, $n=1$ terms.  These can only result from the terms
\beq\label{m0n1}
|\kt\rangle_2\supset \frac{1}{Q_0}\ppin{k_1}\left[ 
\gamma_{2\kt}^{01}(k_1)+\phi_0^2\gamma_{2\kt}^{21}(k_1)
\right]|k_1\rangle_0
\eeq
in $|\kt\rangle_2$.  The last three terms in Eq.~(\ref{schrod}) are then easily written as
\beq
H\p_2|\kt\rangle_2-E_2|\kt\rangle_0-E_1|\kt\rangle_2=\frac{1}{Q_0}\ppin{k_1}\left[ 
(\ok{1}-\okt{})\gamma_{2\kt}^{01}(k_1)-\gamma_{2\kt}^{21}(k_1)
\right]|k_1\rangle_0-\sigma_\kt |\kt\rangle_0.
\eeq
Our strategy will be to determine $\gamma_{2\kt}^{01}(k_1)$ by matching the coefficient of $|k_1\rangle_0$ to
\beq
H\p_4|\kt\rangle_0+H\p_3|\kt\rangle_1=\frac{1}{Q_0}\ppin{k_1}\rho_\kt(k_1)
|k_1\rangle_0+\hat  \sigma_\kt |\kt\rangle_0
\eeq
where $\rho_\kt$ will be calculated below.  Here we have separated out of $\sigma_\kt$ the contribution $\hat\sigma_\kt$ from $\gamma_{2\kt}^{21}$ by decomposing
\beq
\gamma_{2\kt}^{21}(k_1)=\hat\gamma_{2\kt}^{21}(k_1)+2\pi\delta(k_1-\kt)Q_0\left(\hat\sigma_\kt-\sigma_\kt
\right)
\eeq
where $\hat\gamma_\kt(k_1)$ is continuous at $k_1=\kt$.

Matching the coefficients yields
\beq
\gamma_{2\kt}^{01}(k_1)=\frac{-\hat \gamma_{2\kt}^{21}(k_1)+\rho_\kt(k_1)}{\okt{}-\ok{1}}.
\eeq
This is undefined at the two poles, located at $k_1=\pm\kt$.  The ambiguity at $k_1=\kt$ reflects the choice of normalization of the state $|\kt\rangle$.  

The ambiguity at $k_1=-\kt$ results from the fact that the states $|\kt\rangle$ and $|-\kt\rangle$ have the same energy, and both have zero momentum as measured by $P\p$.  We have defined $|\kt\rangle$ as the $H\p$ eigenstate which is $|\kt\rangle_0$ at leading order, however this definition does not fix the mixing with $|-\kt\rangle$ at subleading orders.  We will see below, when we discuss meson multiplication, that a choice of definition of the pole corresponds to a choice of initial condition in meson-kink scattering.  In the future we intend to study elastic kink-meson scattering, with intermediate states consisting of two continuum modes, a continuum mode and a shape mode, or the two shape mode resonance.  We expect that a choice of $i\epsilon$ shift of the pole will be necessary for an initial condition for that process, to ensure that the initial meson is always moving towards the kink.

%we need all terms in $H|\kt\rangle$ that are proportional to $|k_1\rangle$ and are of order $O(\lambda)$.  These are the sum of three kinds of contributions
%\beq
%H|\kt\rangle|_{O(\lambda)}=\left(H\p_4|\kt\rangle_0+H\p_3|\kt\rangle_1+H\p_2|\kt\rangle_2\right)|_{O(\lambda)}.
%\eeq

\subsubsection{Calculating $\rho_\kt$}

Let us begin with $H\p_4|\kt\rangle_0$.  Only one term which appears in Wick's theorem \cite{mewick} will contribute
\bea
H\p_4&=&\frac{\lambda}{24}\int dx \V4 :\phi^4(x):_a\supset \frac{\lambda}{4}\int dx \V4 \left(\I(x) :\phi^2(x):_b+\frac{\I^2(x)}{2}\right)\nonumber\\
&\supset&\frac{\lambda}{2}\ppink{2} V_{\I k_1 -k_2} \Bd 1 \frac{B_{k_2}}{2\ok 2}+\frac{\lambda V_{\I\I}}{8}\hsp
V_{\I\I}=\int dx \V{4} \I^2(x)
\eea
where we have defined the normal ordering $::_b$ which places $B^\ddag$ before $B$.  We then find the contribution
\beq
H\p_4|\kt\rangle_0\supset \frac{\lambda V_{\I\I}}{8}|\kt\rangle_0+\frac{\lambda}{4\okt{}}\ppin{k_1}V_{\I k_1 -\kt} |k_1\rangle_0.
\eeq
The first term contributes to $\hat \sigma_\kt$ and the second to $\rho_\kt$.

Three contributions arise from $H\p_3\ks_1$.  Following Ref.~\cite{menormal}
\bea
H_3\p\ks_1^{00}&=&\frac{\lambda}{8}\left(\frac{V_{\I  -\kt}}{\omega^2_{\kt}}-\frac{\Delta_{-\kt B}}{\omega_\kt\sqrt{\lambda Q_0}}\right)\ppin{k_1}V_{\I k_1}|k_1\rangle_0.\nonumber
\eea
The second is
\bea
H_3\p\ks_1^{02}&=&\frac{\sl}{\sqrt{Q_0}}\ppin{k_1}\left[\ppinkp{2}\frac{\sqrt{\lambda Q_0}V_{-\kt k\p_1 k\p_2}V_{-k\p_1-k\p_2k_1}}{16\omega_\kt\okp1\okp2\left(\omega_\kt-\okp1-\okp2\right)}\right.\nonumber\\
&&\left.+\ppin{k\p}\left(\frac{\left(-\okp{}\Delta_{k\p B}-\sqrt{\lambda Q_0}V_{\I  k\p}\right)
V_{-k\p-\kt k_1}}{8\okp{}^2\omega_\kt}+\frac{\sqrt{\lambda Q_0}V_{-\kt k\p k_1}V_{\I -k\p}}{8\omega_\kt\okp{}\left(\omega_\kt-\okp{}-\ok1\right)}\right)\right.\nonumber\\
&&+\left.
\frac{ \left(-\ok1\Delta_{k_1 B}-\sqrt{\lambda Q_0}V_{\I  k_1}\right)V_{\I -\kt}}{8\omega_\kt\ok1}\right]|k_1\rangle_0\\
&&
+\frac{\sl}{\sqrt{Q_0}}\left[\ppin{k\p}\frac{\left(-\okp{}\Delta_{k\p B}-\sqrt{\lambda Q_0}V_{\I  k\p}\right)
V_{\I -k\p}}{8\okp{}^2}\right]
|\kt\rangle_0.\nonumber
\eea
The third contribution is
\bea
H_3\p\ks_1^{04}&&=-\frac{\lambda}{16}\ppin{k_1}\left[\ppinkp{2}\frac{V_{k_1k\p_1k\p_2}V_{-\kt-k\p_1-k\p_2}}{\omega_\kt\okp1\okp2\left(\ok1+\okp1+\okp2\right)}\right]|k_1\rangle_0\nonumber\\
&&-\frac{\lambda}{48}\left[\ppinkp{3}\frac{V_{k\p_1k\p_2k\p_3}V_{-k\p_1-k\p_2-k\p_3}}{\okp1\okp2\okp3\left(\okp1+\okp2+\okp3\right)}
\right]|\kt\rangle_0.\nonumber
\eea

Adding these together, we may read off
\bea
\hat\sigma_k
&=&\frac{\lambda  V_{\I\I}}{8}+\frac{\sl}{\sqrt{Q_0}}\left[\ppin{k\p}\frac{\left(-\okp{}\Delta_{k\p B}-\sqrt{\lambda Q_0}V_{\I  k\p}\right)
V_{\I -k\p}}{8\okp{}^2}\right]\nonumber\\
&&-\frac{\lambda}{48}\left[\ppinkp{3}\frac{V_{k\p_1k\p_2k\p_3}V_{-k\p_1-k\p_2-k\p_3}}{\okp1\okp2\okp3\left(\okp1+\okp2+\okp3\right)}
\right]
\eea
and
\bea
\rho_\kt(k_1)&=&
\frac{\lambda Q_0}{4\okt{}}V_{\I k_1 -\kt}
+\frac{\lambda Q_0}{8}\left(\frac{V_{\I  -\kt}}{\omega^2_{\kt}}-\frac{\Delta_{-\kt B}}{\omega_\kt\sqrt{\lambda Q_0}}\right)V_{\I k_1}\\
&&+\sqrt{\lambda Q_0}\left[\ppinkp{2}\frac{\sqrt{\lambda Q_0}V_{-\kt k\p_1 k\p_2}V_{-k\p_1-k\p_2k_1}}{16\omega_\kt\okp1\okp2\left(\omega_\kt-\okp1-\okp2\right)}\right.\nonumber\\
&&\left.+\ppin{k\p}\left(\frac{\left(-\okp1\Delta_{k\p B}-\sqrt{\lambda Q_0}V_{\I  k\p}\right)
V_{-k\p-\kt k_1}}{8\okp{}^2\omega_\kt}+\frac{\sqrt{\lambda Q_0}V_{-\kt k\p k_1}V_{\I -k\p}}{8\omega_\kt\okp{}\left(\omega_\kt-\okp{}-\ok1\right)}\right)\right.\nonumber\\
&&+\left.
\frac{ \left(-\ok1\Delta_{k_1 B}-\sqrt{\lambda Q_0}V_{\I  k_1}\right)V_{\I -\kt}}{8\omega_\kt\ok1}\right]
-\frac{\lambda Q_0}{16}\ppinkp{2}\frac{V_{k_1k\p_1k\p_2}V_{-\kt-k\p_1-k\p_2}}{\omega_\kt\okp1\okp2\left(\ok1+\okp1+\okp2\right)}.
\nonumber
\eea

From Ref.~\cite{menormal}
\bea
\gamma_{2\kt}^{21}(k_1)&=&
2\pi\delta(k_1-\kt)\left[\ppin{k\p}\frac{\Delta_{-k\p B}}{8}\left(\Delta_{k\p B}-\frac{\sqrt{\lambda Q_0}V_{\I  k\p}}{\okp{}}\right)\right.\\
&&\left.
-\frac{1}{16}\ppinkp{2}\frac{\left(\okp1-\okp2\right)^2}{\okp1\okp2}\Delta_{k\p_1k\p_2}\Delta_{-k\p_1,-k\p_2}\right]\nonumber\\
&&+\frac{3}{8}\left(-1+\frac{\ok1}{\omega_{\kt}}\right)\Delta_{k_1 B}\Delta_{-\kt B}-\frac{1}{4}\ppin{k\p}\left(\frac{\ok1}{\okp{}}+\frac{\okp{}}{\omega_{\kt}}
\right)\Delta_{-\kt,-k\p}\Delta_{k_1k\p}\nonumber\\
&&-\frac{\sqrt{\lambda Q_0}}{8\omega_\kt}\left(\omega_{k_1}\Delta_{k_1 B}\frac{V_{\I -\kt}}{\omega_\kt}+\omega_{\kt}\Delta_{-\kt B}\frac{V_{\I  k_1}}{\ok1}\right)+\frac{1}{8}\ppin{k\p}
\frac{\sqrt{\lambda Q_0}\Delta_{-k\p B}V_{-\kt k_1 k\p}}{\omega_\kt\left(\omega_\kt-\ok1-\okp{}\right)}.\nonumber
\eea
Decomposing, this is
\bea
\hat\gamma_{2\kt}^{21}(k_1)&=&\frac{3}{8}\left(-1+\frac{\ok1}{\omega_{\kt}}\right)\Delta_{k_1 B}\Delta_{-\kt B}-\frac{1}{4}\ppin{k\p}\left(\frac{\ok1}{\okp{}}+\frac{\okp{}}{\omega_{\kt}}
\right)\Delta_{-\kt,-k\p}\Delta_{k_1k\p}\nonumber\\
&&-\frac{\sqrt{\lambda Q_0}}{8\omega_\kt}\left(\omega_{k_1}\Delta_{k_1 B}\frac{V_{\I -\kt}}{\omega_\kt}+\omega_{\kt}\Delta_{-\kt B}\frac{V_{\I  k_1}}{\ok1}\right)+\frac{1}{8}\ppin{k\p}
\frac{\sqrt{\lambda Q_0}\Delta_{-k\p B}V_{-\kt k_1 k\p}}{\omega_\kt\left(\omega_\kt-\ok1-\okp{}\right)}\nonumber\\
\hat\sigma_\kt-\sigma_\kt&=&\frac{1}{Q_0}\left[\ppin{k\p}\frac{\Delta_{-k\p B}}{8}\left(\Delta_{k\p B}-\frac{\sqrt{\lambda Q_0}V_{\I  k\p}}{\okp{}}\right)\right.\nonumber\\
&&\left.
-\frac{1}{16}\ppinkp{2}\frac{\left(\okp1-\okp2\right)^2}{\okp1\okp2}\Delta_{k\p_1k\p_2}\Delta_{-k\p_1,-k\p_2}\right].
\eea

In particular this implies $\sigma_\kt=Q_2$, where $Q_2$ is the two-loop correction to the kink ground state mass, found in Ref.~\cite{me2loop}.

\subsubsection{The Reduced Inner Products}

The inner product of the $O(\lambda)$ correction $|\kt\rangle_2$ with the leading term $|\kt\rangle_0$ yields
\bea
\langle\kt_1|\kt_2\rangle_{\rm{red}}&\supset&
\frac{1}{\sqrt{Q_0}}\frac{\gamma^{01}_{2\kt_2}(\kt_1)}{2\okt{1}}+\frac{1}{\sqrt{Q_0}}\frac{\gamma^{01*}_{2\kt_1}(\kt_2)}{2\okt{2}}\\
%&=&\frac{\hat \gamma_{2\kt_2}^{21}(\kt_1)+\frac{\rho_{\kt_2}(\kt_1)}{\okt{2}-\okt{1}}}{2\okt 1}
%+\frac{\hat \gamma_{2\kt_1}^{21}(\kt_2)+\frac{\rho_{\kt_1}(\kt_2)}{\okt{1}-\okt{2}}}{2\okt 2}.
%\nonumber\\
&=&\frac{-\okt2 \hat \gamma_{2\kt_2}^{21}(\kt_1)+\okt1 \hat \gamma_{2\kt_1}^{21 *}(\kt_2)}{2\sqrt{Q_0}\okt 1\okt 2(\okt 2-\okt 1)}
+\frac{\okt 2\rho_{\kt_2}(\kt_1)-\okt 1\rho^*_{\kt_1}(\kt_2)}{2\sqrt{Q_0}\okt 1\okt 2(\okt 2-\okt 1)}.
\nonumber
\eea
Due to the antisymmetry, there are many cancellations in the second numerator
\bea
&&\okt 2\rho_{\kt_2}(\kt_1)=
\frac{\lambda Q_0}{4}V_{\I \kt_1 -\kt_2}
+\frac{\lambda Q_0}{8}\left(\frac{V_{\I  -\kt_2}}{\omega_{\kt_2}}-\frac{\Delta_{-\kt_2 B}}{\sqrt{\lambda Q_0}}\right)V_{\I \kt_1}\nonumber\\
&&+\frac{\lambda Q_0}{16}\ppinkp{2}\frac{V_{-\kt_2 k\p_1 k\p_2}V_{-k\p_1-k\p_2\kt_1}}{\okp1\okp2\left(\okt2-\okp1-\okp2\right)}\nonumber\\
&&+\frac{\sqrt{\lambda Q_0}}{8}\ppin{k\p}\left(\frac{\left(-\okp1\Delta_{k\p B}-\sqrt{\lambda Q_0}V_{\I  k\p}\right)
V_{-k\p-\kt_2 \kt_1}}{\okp{}^2}+\frac{\sqrt{\lambda Q_0}V_{-\kt_2 k\p \kt_1}V_{\I -k\p}}{\okp{}\left(\okt2-\okt1-\okp{}\right)}\right)\nonumber\\
&&+
\frac{\lambda Q_0}{8} \left(-\frac{\Delta_{k_1 B}}{\sqrt{\lambda Q_0}}-\frac{V_{\I  \kt_1}}{\okt1}\right)V_{\I -\kt_2}
-\frac{\lambda Q_0}{16}\ppinkp{2}\frac{V_{\kt_1k\p_1k\p_2}V_{-\kt_2-k\p_1-k\p_2}}{\okp1\okp2\left(\okt{1}+\okp1+\okp2\right)}
\eea
and so
\bea
\frac{\okt 2\rho_{\kt_2}(\kt_1)-\okt 1\rho^*_{\kt_1}(\kt_2)}{2\sqrt{Q_0}\okt1\okt2(\okt2-\okt1)} \label{diff}
&&=-\frac{\lambda \sqrt{Q_0}}{8}\frac{V_{\I-\kt_2}V_{\I\kt_1}}{\okt1^2\okt2^2}\\
&&-\frac{\lambda \sqrt{Q_0}}{32\okt1\okt2}\ppinkp{2}\frac{V_{-\kt_2 k\p_1 k\p_2}V_{-k\p_1-k\p_2\kt_1}}{\okp1\okp2\left(\okt2-\okp1-\okp2\right)\left(\okt1-\okp1-\okp2\right)}
\nonumber\\
&&+\frac{\lambda \sqrt{Q_0}}{8\okt1\okt2}\ppin{k\p}\frac{V_{-\kt_2 k\p \kt_1}V_{\I -k\p}}{\okp{}\left[(\okt2-\okt1)^2-\okp{}^2\right]}\nonumber\\
&&-\frac{\lambda \sqrt{Q_0}}{32\okt1\okt2}\ppinkp{2}\frac{V_{\kt_1k\p_1k\p_2}V_{-\kt_2-k\p_1-k\p_2}}{\okp1\okp2\left(\okt{1}+\okp1+\okp2\right)\left(\okt{2}+\okp1+\okp2\right)}.\nonumber
\eea
Similarly
\bea
\okt2 \hat \gamma_{2\kt_2}^{21}(\kt_1)&=&
\frac{3}{8}\left({\okt1}-\okt2\right)\Delta_{\kt_1 B}\Delta_{-\kt_2 B}-\frac{1}{4}\ppin{k\p}\left(\frac{\okt1\okt2}{\okp{}}+{\okp{}}{}
\right)\Delta_{-\kt_2,-k\p}\Delta_{\kt_1k\p}\nonumber\\
&&-\frac{\sqrt{\lambda Q_0}}{8}\left(\okt1\Delta_{\kt_1 B}\frac{V_{\I -\kt_2}}{\okt2}+\okt2\Delta_{-\kt_2 B}\frac{V_{\I  \kt_1}}{\okt1}\right)\nonumber\\
&&+\frac{1}{8}\ppin{k\p}
\frac{\sqrt{\lambda Q_0}\Delta_{-k\p B}V_{-\kt_2 \kt_1 k\p}}{\left(\okt2-\okt1-\okp{}\right)} \label{somma}
\eea
leading to
\bea
\frac{-\okt2 \hat \gamma_{2\kt_2}^{21}(\kt_1)+\okt1 \hat \gamma_{2\kt_1}^{21 *}(\kt_2)}{2\sqrt{Q_0}\okt1\okt2\left({\okt2}-\okt1\right)}&=&\frac{3\Delta_{\kt_1 B}\Delta_{-\kt_2 B}}{8\sqrt{Q_0}\okt1\okt2}-\frac{1}{8\sqrt{Q_0}\okt1\okt2}\ppin{k\p}
\frac{\sqrt{\lambda Q_0}\Delta_{-k\p B}V_{-\kt_2 \kt_1 k\p}}{\left[(\okt2-\okt1)^2-\okp{}^2\right]} .\label{gh}
\nonumber
\eea
%\bea
%\frac{\okt2 \hat \gamma_{2\kt_2}^{21}(\kt_1)+\okt1 \hat \gamma_{2\kt_1}^{21}(\kt_2)}{2\okt1\okt2}&=&-\frac{1}{4}\ppin{k\p}\left(\frac{1}{\okp{}}+\frac{\okp{}}{\okt1\okt2}
%\right)\Delta_{-\kt_2,-k\p}\Delta_{\kt_1k\p}\nonumber\\
%&&-\frac{\sqrt{\lambda Q_0}}{8}\left(\Delta_{\kt_1 B}\frac{V_{\I -\kt_2}}{\okt2^2}+\Delta_{-\kt_2 B}\frac{V_{\I  \kt_1}}{\okt1^2}\right)\nonumber\\
%&&+\frac{\sqrt{\lambda Q_0}}{2\okt1\okt2}\ppin{k\p}
%\frac{\okp{}\Delta_{-k\p B}V_{-\kt_2 \kt_1 k\p}}{4\left[(\okt2-\okt1)^2-\okp{}^2\right]}.
%\eea
This concludes our calculation of both contributions (\ref{diff}) and (\ref{gh}) to the reduced inner product $\langle\kt_1|\kt_2\rangle_{\rm{red}}$ involving the $O(\lambda)$ corrections to the states, encoded in $\gamma_2$.  

We will now calculate contributions to the inner product involving only terms of $O(\lambda^0)$ and $O(\sl)$, encoded in $\gamma_0$ and $\gamma_1$ respectively.  We will define $\langle\kt_1|\kt_2\rangle_{n,\rm{red}}$ to consist of all such terms in Eq.~(\ref{padqft}) at the corresponding value of $n$.  Then, using the coefficients in Eqs.~(\ref{g0}) and (\ref{gammakt}), one finds
\bea
\langle\kt_1|\kt_2\rangle_{\rm{1,red}}&=&
\ppin{k_1} \frac{\gamma_{\kt_1}^{01*}(k_1)}{ (2\ok{1})}\left[\sqrt{Q_0}\gamma_{\kt_2}^{01}(k_1)+\Delta_{k_1 B}\gamma_{\kt_2}^{00}+2\ppin{k\p}\frac{\Delta_{k\p B}}{2\okp{}}{\gamma_{\kt_2}^{02}(-k\p,k_1)}
\right]\nonumber\\
&=& \frac{1}{ 2\okt{1}}\left[\sqrt{Q_0}\gamma_{\kt_2}^{01}(\kt_1)+\Delta_{\kt_1 B}\gamma_{\kt_2}^{00}+2\ppin{k\p}\frac{\Delta_{k\p B}}{2\okp{}}{\gamma_{\kt_2}^{02}(-k\p,\kt_1)}
\right]\nonumber\\
&=&\frac{\sqrt{Q_0}2\pi\delta(\kt_1-\kt_2)}{ 2\okt{1}}+\frac{1}{ 2\okt{1}\sqrt{Q_0}}\left[ 
\Delta_{\kt_1 B}\left(
 \frac{\sqrt{Q_0\lambda}V_{\I -\kt_2}}{4\okt{2}^2}-\frac{\Delta_{-\kt_2 B}}{4\okt 2}
\right)
\right.\nonumber\\
&&\left.+\ppin{k\p}\frac{\Delta_{k\p B}}{\okp{}}\left(
 -\frac{2\pi\delta(\kt_1-\kt_2)}{4}\left(\Delta_{-k\p B}+\sqrt{Q_0\lambda}\frac{V_{\I  -k\p}}{\okp{}}\right)\right.\right.\nonumber\\
 &&\left.\left.+\frac{\sqrt{Q_0\lambda}V_{-\kt_2 -k\p \kt_1}}{4\okt 2\left(\okt 2-\okp{}-\okt 1\right)}-\frac{2\pi\delta(k\p+\kt_2)}{4}\left(\Delta_{\kt_1 B}+\sqrt{Q_0\lambda}\frac{V_{\I  \kt_1}}{\okt 1}\right)
\right)
\right]\nonumber\\
&=&\frac{2\pi\delta(\kt_1-\kt_2)}{2\okt 1}\left[ 
{\sqrt{Q_0}}
-\frac{1}{\sqrt{Q_0}}\ppin{k\p}\frac{\Delta_{k\p B}}{4\okp{}}\left(\Delta_{-k\p B}+\sqrt{Q_0\lambda}\frac{V_{\I  -k\p}}{\okp{}}\right)
\right]
\nonumber\\
&&
+\frac{\Delta_{\kt_1 B}}{ 8\okt{1}\okt 2\sqrt{Q_0}}
\left(
 \frac{\sqrt{Q_0\lambda}V_{\I -\kt_2}}{\okt{2}}-{\Delta_{-\kt_2 B}}{}
\right)
-\frac{\Delta_{-\kt_2 B}}{8\okt 1\okt{2}\sqrt{Q_0}}\left({\Delta_{\kt_1 B}}{}+\sqrt{Q_0\lambda}\frac{V_{\I  \kt_1}}{\okt 1}\right)
\nonumber\\
&&+\frac{\sqrt{\lambda}}{8\okt1\okt2}\ppin{k\p}\frac{\Delta_{k\p B}V_{-\kt_2 -k\p \kt_1}}{\okp{}(\okt 2-\okp{}-\okt 1)}
\label{eq1}
\eea
and
\bea
\langle\kt_1|\kt_2\rangle_{\rm{0,red}}&=&\frac{1}{16\sqrt{Q_0}\omega_{\kt_1}\omega_{\kt_2}}\left(\frac{\sqrt{Q_0\lambda}V_{\I \kt_1}}{\omega_{\kt_1}}-\Delta_{\kt _1B}\right)\left(\frac{\sqrt{Q_0\lambda}V_{\I -\kt_2}}{\omega_{\kt_2}}+\Delta_{-\kt_2B}\right)
\label{eq0}
\eea
and
\bea
\langle\kt_1|\kt_2\rangle_{\rm{2,red}}&=&\ppink{2} \frac{\gamma_{\kt_1}^{02*}(k_1,k_2)}{4\ok 1\ok 2}\left[\sqrt{Q_0}\gamma_{\kt_2}^{02}(k_1,k_2)+\Delta_{k_1 B}\gamma_{\kt_2}^{01}(k_2)+(k_1\leftrightarrow k_2)\right]\nonumber\\
&=&\ppink{2} \frac{1}{16\ok 1\ok 2\sqrt{Q_0}}\left[-{2\pi\delta(k_2-\kt_1)}\left(\Delta_{-k_1 B}+\sqrt{Q_0\lambda}\frac{V_{\I  -k_1}}{\ok1}\right)\right.\nonumber\\
&&\left.+\frac{\sqrt{Q_0\lambda}V^*_{-\kt_1 k_1 k_2}}{\omega_{\kt_1}\left(\omega_{\kt_1}-\ok1-\ok2\right)}-{2\pi\delta(k_1-\kt_1)}{}\left(\Delta_{-k_2 B}+\sqrt{Q_0\lambda}\frac{V_{\I  -k_2}}{\ok 2}\right)\right]\nonumber\\
&&\times\left[ {2\pi\delta(k_2-\kt_2)}{}\left(\Delta_{k_1 B}-\sqrt{Q_0\lambda}\frac{V_{\I  k_1}}{\ok1}\right)+\frac{\sqrt{Q_0\lambda}V_{-\kt_2 k_1 k_2}}{2\omega_{\kt_2}\left(\omega_{\kt_2}-\ok1-\ok2\right)}\right]\nonumber\\
&=&\frac{2\pi\delta(\kt_1-\kt_2)}{16\omega_{\kt_1}\sqrt{Q_0}}\pin{k_1}\frac{1}{\ok 1}\left[Q_0\lambda\frac{|V_{\I k_1}|^2}{\ok{1}^2}-|\Delta_{k_1B}|^2  
\right]\nonumber\\
&&+\frac{1}{16\omega_{\kt_1}\omega_{\kt_2}\sqrt{Q_0}}
\left(\sqrt{Q_0\lambda}\frac{V_{\I  -\kt_2}}{\omega_{\kt_2}}+\Delta_{-\kt_2 B}\right)\left(\sqrt{Q_0\lambda}\frac{V_{\I  \kt_1}}{\omega_{\kt_1}}-\Delta_{\kt_1 B}\right)
\nonumber\\
&&+\frac{\sqrt{\lambda}}{16\omega_{\kt_1}\omega_{\kt_2}}\ppin{k\p}\frac{V_{\kt_1-\kt_2 -k\p }}{\okp{}\left(\omega_{\kt_1}-\omega_{\kt_2}-\okp{}\right)}\left(\Delta_{k\p B}-\sqrt{Q_0\lambda}\frac{V_{\I  k\p}}{\okp{}}\right)\nonumber\\
&&+\frac{\sqrt{\lambda}}{16\omega_{\kt_1}\omega_{\kt_2}}\ppin{k\p}\frac{V_{\kt_1-\kt_2 -k\p }}{\okp{}\left(\omega_{\kt_1}-\omega_{\kt_2}+\okp{}\right)}\left(\Delta_{k\p B}+\sqrt{Q_0\lambda}\frac{V_{\I  k\p}}{\okp{}}\right)
\nonumber\\
&&+\frac{\sqrt{Q_0}\lambda}{32\omega_{\kt_1}\omega_{\kt_2}}\ppinkp{2}\frac{V_{\kt_1 -k\p_1 -k\p_2}V_{-\kt_2 k\p_1 k\p_2}}{\okp1\okp2\left(\omega_{\kt_1}-\okp1-\okp2\right)\left(\omega_{\kt_2}-\okp1-\okp2\right)}.
\eea
The $V_{\I -\kt_2}V_{\I\kt_1}$ term on the second line, plus that in Eq.~(\ref{eq0}) cancel that in the first line of Eq.~(\ref{diff}).  The $\Delta_{-\kt_2 B}\Delta_{\kt_1 B}$ term on the second line, added to the contribution in Eq.~(\ref{eq0}) and the two contributions in Eq.~(\ref{eq1}) exactly cancels the first term in Eq.~(\ref{gh}).  Adding the $V_{\kt_1-\kt_2-k\p}\Delta_{k\p B}$ terms on the third and fourth lines to the last line of Eq.~(\ref{eq1}) leads to a total which precisely cancels the other term in Eq.~(\ref{gh}).

Adding the third and forth lines, the $V_{\kt_1-\kt_2 k\p}V_{\I k\p}$ term cancels that on the third line of Eq.~(\ref{diff}).  The last line cancels the second line of Eq.~(\ref{diff}).  Finally, the $\Delta_{\kt_1}V_{\I\kt_2}$ terms in the second line, added to that in Eq.~(\ref{eq0}) and in the second line of the last expression of Eq.~(\ref{eq1}) vanishes, as does its conjugate $(\kt_1\leftrightarrow \kt_2)$.  

Finally, the $n=4$ contribution is
\bea
\langle\kt_1|\kt_2\rangle_{\rm{4,red}}&=&2\pi\delta(\kt_1-\kt_2)\frac{\lambda\sqrt{Q_0}}{96\omega_{\kt_1}}\ppink{3}
\frac{|V_{k_1k_2k_3}|^2}{\ok{1}\ok{2}\ok{3}(\ok{1}+\ok{2}+\ok{3})^2}\nonumber\\
&&+\frac{\lambda\sqrt{Q_0}}{32
\omega_{\kt_1}\omega_{\kt_2}}\ppink{2}
\frac{V^*_{\kt_2 k_1 k_2}V_{\kt_1 k_1 k_2}}{\ok{1}\ok{2}(\okt 1+\ok{1}+\ok{2})(\okt 2+\ok{1}+\ok{2})}.
\eea
The second line, which is the only one which survives at $\kt_1\neq\kt_2$, cancels the last line of Eq.~(\ref{diff}), completing the cancellation of the terms in Eq.~(\ref{diff}).

\subsubsection{Remarks}

Thus we conclude that at $\kt_1\neq\kt_2$ the reduced inner product vanishes.  This is as it must be, as these represent distinct eigenstates of $H\p$.  It is thus a consistency test of our main result (\ref{padqft}).

Our derivation does not apply to $O(\sl)$ corrections at $\kt_1=-\kt_2$ as there have been terms with $(\okt1-\okt2)$ in both the numerator and denominator, which we have canceled.  Indeed the states $|\kt_1\rangle$ and $|-\kt_1\rangle$ have the same energy and so may mix.  

The problem is not simply that we were not careful, indeed there is a degenerate eigenspace  and so one is free to define $|\kt_1\rangle$ to have any overlap with $|-\kt_1\rangle$.  However, for a given physical problem, there may be a more useful prescription for the pole at $\okt1=\okt2$.  When we turn to meson multiplication below, we will see how such a physical principle fixes a related pole.  In future work, we intend to use elastic meson-kink scattering to fix the prescription for defining the pole at $\kt_1=-\kt_2.$

Similarly, our derivation is not reliable at $\kt_1=\kt_2$ as the same manipulation is ill-defined.  This is simply a reflection of our freedom to choose the normalization of $|\kt_1\rangle$.   One may, for example, fix $\gamma_{i\kt}^{01}(\kt)=0$ for all $i>0$, analogously to the condition $\gamma_2^{00}=0$ that we imposed when computing the reduced norm of the 1-kink, 0-meson state.

\section{Initial and Final State Corrections} \label{multsez}

\subsection{Motivation}

In an experiment, any initial condition is allowed.  The choice of initial condition is at the discretion of the experimenter, as it depends on how the experiment is set up.  Similarly, the choice of final states in each detection channel is determined by the experimenter, as it depends on the design of the detector.  In Ref.~\cite{memult} we considered initial state wave packets constructed as a superposition of $H\p_2$ eigenstates $|k_1\rangle_0$, each corresponding to the first semiclassical approximation of the desired state.  In other words, the initial one-meson state was constructed exclusively using the one-meson Fock space of the free kink Hamiltonian $H\p_2$.  Similarly, the probability calculated used a projector onto the two-meson Fock space of the free Hamiltonian, which is generated by the states $|k_2k_3\rangle_0$.  This procedure is well-defined and corresponds to the result of some experiment.

However, there was a choice.  One could, instead, have used eigenstates $|k_1\rangle$ of the full Hamiltonian $H\p$ to build the wave packet.  Each element of the one-meson Fock space of the full Hamiltonian $H\p$ contains a superposition of the various $n$-meson eigenstates $|k_1\cdots k_n\rangle_0$ of the free Hamiltonian $H\p_2$.  This choice is somewhat arbitrary as the wave packet itself will not be the eigenstate of either Hamiltonian.  However, one may ask whether the resulting probability depends on this choice.  This is an important point experimentally because, if the probability depends on the choice, then one needs to determine just to which choice a given preparation method and detector corresponds.  Theoretically it is also important because, if the results differ, one choice may be compatible with an LSZ reduction theorem while the other may not.

\subsection{The Initial and Final Conditions}

In Ref.~\cite{memult} we calculated the amplitude for an initial state with one kink and one meson to evolve to a final state with one kink and two mesons.  We called this process meson multiplication.  While the initial state and final state involved no powers of $\lambda$, the interaction contained a $\sqrt{\lambda}$ and so the amplitude was of order $O(\sqrt{\lambda})$.  However, if the initial state contained a quantum correction of $O(\sqrt{\lambda})$ which could evolve via the $\lambda$-free $H\p_2$ to the final state, this would contribute at the same order.  Similarly, if the admissible final states contained an $O(\sqrt{\lambda})$  correction which has an $O(1)$ inner product with the $H\p_2$-evolved initial state, it will also contribute at the same order.  Just such corrections arise if our initial state or projector is constructed as superpositions of eigenstates of the full Hamiltonian.

Thus we are motivated to consider a reflectionless kink, so that far from the kink the normal modes become plane waves, whose form we will review shortly in Eq.~(\ref{gk}). Letting the initial meson wave packet have the same superposition coefficients as in Ref.~\cite{memult}
\beq
\alpha_{k_1}=2\sigma\sqrt{\pi}\mb_{k_1}e^{-\sigma^2\left(k_1-k_0\right)^2}e^{i(k_0-k_1)x_0}
\eeq
but this time, as a superposition of the 1-meson states $|k_1\rangle$ which are eigenstates of the full kink Hamiltonian $H\p$.  Our initial state is
\beq
\left|\Phi\right\rangle=\int \frac{d k_1}{2 \pi} \alpha_{k_1}\left|k_1\right\rangle \label{is}
\eeq
unlike Ref.~\cite{memult} where the $H\p$-eigenstate $|k_1\rangle$ was replaced with, in the notation of the present paper, the $H\p_2$-eigenstate $|k_1\rangle_0$
\beq
\left|\Phi\right\rangle_0=\int \frac{d k_1}{2 \pi} \alpha_{k_1}\left|k_1\right\rangle_0.
\eeq
Note that in both cases, one integrates over continuum modes $k_1$ with no sum over bound modes, as these vanish exponentially far from the kink, and we have assumed that $|x_0|\gg 1/m$.

Now, instead of the matrix element ${}_0\langle k_2 k_3|e^{-it(H\p_2+H\p_3)}|\Phi\rangle_0$ computed in Ref.~\cite{memult}, we will be interested in a matrix element which we write as
\beq
{}_\rv\langle k_2 k_3|e^{-itH\p}|\Phi\rangle.
\eeq
Here $|k_2k_3\rangle_\rv$ is not the kink Hamiltonian eigenstate $|k_2k_3\rangle$.  If it were, then the $H\p$ in the evolution operator would just multiply it by a phase and the matrix element would evolve by a simple phase rotation, whereas we are interested in a probability which begins at zero and evolves to a nonzero value.  Rather it is the translation-invariant eigenstate of $H\p$ far to the left or right of the kink, which is defined by replacing $f(x)$ with $f(-\infty)$ or $f(+\infty)$ in its definition (\ref{dfd},\ref{df}).  

In the case of a nonreflective kink, at leading order, the only relevant quantum correction in $|k_2k_3\rangle_\rv$ is
\beq
|k_2k_3\rangle_\rv=|k_2k_3\rangle_0+\frac{\sl V^{(3)}(\sl f(-\infty))\mb_{-k_2}\mb_{-k_3}\mb_{k_2+k_3}}{4\ok2\ok3(\ok2+\ok3-\omega_{k_2+k_3})}|k_2+k_3\rangle_0\label{sf}
\eeq
for an inner product with a wave packet localized at $x\ll 0$ and
\beq
|k_2k_3\rangle_\rv=|k_2k_3\rangle_0+\frac{\sl V^{(3)}(\sl f(+\infty))\md_{-k_2}\md_{-k_3}\md_{k_2+k_3}}{4\ok2\ok3(\ok2+\ok3-\omega_{k_2+k_3})}|k_2+k_3\rangle_0\label{sf2}
\eeq
for an inner product with a wave packet localized at $x\gg 0$.  The projector $\mathcal{P}$ is assembled from an integral of wave packets of $|k_2k_3\rangle_{\rm{vac}}$ localized at $x\ll 0$ and $x\gg 0$ consisting of superpositions of (\ref{sf}) and (\ref{sf2}) respectively.  Note that only the first term in  (\ref{sf}) and in (\ref{sf2}) will be relevant to initial state corrections, and the second to final state corrections.

\subsection{Initial State Corrections} \label{isc}

The state at time $t$ is
\beq
|t\rangle=\pin{k_1}\alpha_{k_1}
  e^{-itH\p}|k_1\rangle=\pin{k_1}\alpha_{k_1}
  e^{-it\tilde{\omega}_{k_1}}|k_1\rangle
\eeq
where $\tilde{\omega}_{k_1}$ is the quantum corrected energy of $|k_1\rangle$.  It is equal to $\ok{1}$ plus corrections of order $O(\lambda)$ \cite{menormal}.  As we are only considering corrections of order $O(\sqrt{\lambda})$ here, we can ignore these corrections and set it to $\ok{1}$.  Next, as $\alpha_{k_1}$ is localized near $k_1=k_0$, we may expand
\beq
\tilde{\omega}_{k_1}=\ok{1}=\ok{0}+\frac{k_0}{\ok{0}}(k_1-k_0).
\eeq
We then find
\bea
|t\rangle&=&\pin{k_1}2\sigma\sqrt{\pi}\mb_{k_1}e^{-\sigma^2\left(k_1-k_0\right)^2}e^{i(k_0-k_1)x_0}
  e^{-it\left(\ok{0}+\frac{k_0}{\ok{0}}(k_1-k_0)\right)}|k_1\rangle\\
&=&2\sigma\sqrt{\pi}\mb_{k_0}e^{-i\ok{0}t}%\rm{Exp}\left[-\frac{\left(x_0+\frac{k_0}{\ok{0}}t\right)^2}{4\sigma^2}\right]
\pin{k_1}e^{-\sigma^2\left(k_1-k_0\right)^2}e^{-i(k_1-k_0)\left(x_0+\frac{k_0}{\ok{0}}t\right)}|k_1\rangle.
\nonumber
\eea

Now, let us a consider a specific contribution to $|k_1\rangle$ at $O(\sqrt{\lambda})$
\bea
|k_1\rangle&\supset&\frac{1}{\sqrt{Q_0}}\pin{k_2}\pin{k_3}\gamma_{1k_1}^{02}(k_2,k_3) |k_2k_3\rangle_0 \label{spec}\\
&\supset&
\frac{\sqrt{\lambda}}{4\ok 1}\pin{k_2}\pin{k_3}\frac{V_{-k_1 k_2 k_3}}{\left(\ok1-\ok2-\ok3\right)} |k_2k_3\rangle_0.
\nonumber
\eea
The case in which $k_2$ or $k_3$ is a bound mode is interesting and will be the subject of a separate study on (anti)Stokes scattering, and so here we will consider only continuum modes $k_2$ and $k_3$.  There is a pole at $\ok{1}=\ok{2}+\ok{3}$.  This pole of course is important, as meson multiplication occurs on the pole.  But let us first consider $k_2$ and $k_3$ far from this pole, as compared with $1/\sigma$, returning to the pole in Subsec.~\ref{polesez}.  Then we can set $k_1$ to $k_0$ in the denominator and (\ref{spec}) contributes
\bea
|t\rangle&\supset&\frac{\sqrt{\lambda}\mb_{k_0}e^{-i\ok{0}t}}{4\ok 0}\pin{k_2}\pin{k_3}\frac{1}{\ok 0-\ok 2-\ok 3}\int dx \V3 \g_{k_2}(x)\g_{k_3}(x)\nonumber\\
&&\times  \left[2\sigma\sqrt{\pi}\pin{k_1} e^{-\sigma^2\left(k_1-k_0\right)^2}e^{-i(k_1-k_0)\left(x_0+\frac{k_0}{\ok{0}}t\right)}\g_{-k_1}(x)\right]|k_2 k_3\rangle_0. \label{speccon}
\eea
Let us try to evaluate the integral in square brackets at $x\gg 0$ and $x\ll 0$, where
\bea
\g_k(x)&=&\left\{\begin{tabular}{lll}
$\mb_ke^{-ikx}$&\rm{if} & $x\ll  -1/m$\\
$\md_ke^{-ikx}$&\rm{if} & $x\gg 1/m$\\
\end{tabular}
\right. \label{gk}\\
|\mb_k|^2&=&|\md_k|^2=1\hsp
\mb^*_k=\mb_{-k}\hsp
\md^*_k=\md_{-k}.\nonumber
\eea
It is
\bea
&&2\sigma\sqrt{\pi}\pin{k_1} e^{-\sigma^2\left(k_1-k_0\right)^2}e^{-i(k_1-k_0)\left(x_0+\frac{k_0}{\ok{0}}t\right)}\g_{-k_1}(x)\\
&&\hspace{3cm}=e^{ik_0 x}{\rm{Exp}}\left[-\frac{\left(-x+x_0+\frac{k_0}{\ok 0}t\right)^2}{4\sigma^2}\right]\left\{\begin{tabular}{lll}
$\mb_{-k_0}$&\rm{if} & $x\ll  -1/m$\\
$\md_{-k_0}$&\rm{if} & $x\gg 1/m$.\\
\end{tabular}
\right. \nonumber
\eea

We see that $x$  is peaked near $x_t$ where
\beq
x_t=x_0+\frac{k_0}{\ok 0}t.
\eeq
When $|x_t|\gg 0$, the Gaussian is supported at $|x|\gg 0$.  Here $f(x)$ tends to a constant, and so $\V3$ also tends to a constant, corresponding to the third derivative of the potential in one of the two vacua of the theory.  The value of the constant depends on the sign of $x_t$.  Now let us turn to the $x$ integration.  For concreteness, let us consider $t$ much smaller than the time when the meson wave packet strikes the kink, so that $x\ll 0$, then
\bea
&&\int dx \V3 \g_{k_2}(x)\g_{k_3}(x) e^{ik_0 x}{\rm{Exp}}\left[-\frac{\left(-x+x_0+\frac{k_0}{\ok 0}t\right)^2}{4\sigma^2}\right]\mb_{-k_0}\\
&&\hspace{2cm}=2\sigma\sqrt{\pi}V^{(3)}(\sqrt{\lambda}f(-\infty))\mb_{-k_0}\mb_{k_2}\mb_{k_3}e^{-\sigma^2(k_0-k_2-k_3)^2}e^{ix_t(k_0-k_2-k_3)}.
\nonumber
\eea
When $t$ is large, so that $x_t\gg 0$, one simply changes the phases $\mb$ into $\md$ and $V^{(3)}$ is evaluated at the vacuum on the right of the kink.  

Summarizing, after the collision
\bea
|t\rangle&\supset&\frac{2\sigma\sqrt{\pi}V^{(3)}(\sqrt{\lambda}f(+\infty))\sqrt{\lambda}\mb_{k_0}\md_{-k_0}e^{-i\ok{0}t}}{4\ok 0}\label{prima}\\
&&\times\pin{k_2}\pin{k_3}\frac{\md_{k_2}\md_{k_3}}{\ok 0-\ok 2-\ok 3}e^{-\sigma^2(k_0-k_2-k_3)^2}e^{ix_t(k_0-k_2-k_3)} |k_2 k_3\rangle_0\nonumber\\
&=&\frac{2\sigma\sqrt{\pi}V^{(3)}(\sqrt{\lambda}f(+\infty))\sqrt{\lambda}\mb_{k_0}\md_{-k_0}e^{-i\ok{0}t+ik_0x_t}}{4\ok 0}\nonumber\\
&&\times\pin{k_2}\pin{k_3}\frac{\g_{k_2}(x_t)\g_{k_3}(x_t)}{\ok 0-\ok 2-\ok 3}e^{-\sigma^2(k_0-k_2-k_3)^2} |k_2 k_3\rangle_0\nonumber\\
&=&\frac{V^{(3)}(\sqrt{\lambda}f(+\infty))\sqrt{\lambda}\mb_{k_0}\md_{-k_0}e^{-i\ok{0}t}}{4\ok 0}\nonumber\\
&&\times\pin{k_2}\pin{k_3}\frac{\md_{k_2}\md_{k_3}}{\ok 0-\ok 2-\ok 3}2\pi\delta(k_0-k_2-k_3) |k_2 k_3\rangle_0.\nonumber
\eea
In the last equality we considered the limit $\sigma\rightarrow\infty$.  Before the collision
\bea
|t\rangle&\supset&\frac{2\sigma\sqrt{\pi}V^{(3)}(\sqrt{\lambda}f(-\infty))\sqrt{\lambda}\mb_{k_0}\mb_{-k_0}e^{-i\ok{0}t}}{4\ok 0}\label{dopo}\\
&&\times\pin{k_2}\pin{k_3}\frac{\mb_{k_2}\mb_{k_3}}{\ok 0-\ok 2-\ok 3}e^{-\sigma^2(k_0-k_2-k_3)^2}e^{ix_t(k_0-k_2-k_3)} |k_2 k_3\rangle_0.\nonumber\\
&=&\frac{2\sigma\sqrt{\pi}V^{(3)}(\sqrt{\lambda}f(-\infty))\sqrt{\lambda}e^{-i\ok{0}t+ik_0x_t}}{4\ok 0}\nonumber\\
&&\times\pin{k_2}\pin{k_3}\frac{\g_{k_2}(x_t)\g_{k_3}(x_t)}{\ok 0-\ok 2-\ok 3}e^{-\sigma^2(k_0-k_2-k_3)^2} |k_2 k_3\rangle_0\nonumber\\
&=&\frac{V^{(3)}(\sqrt{\lambda}f(-\infty))\sqrt{\lambda}e^{-i\ok{0}t}}{4\ok 0}\nonumber\\
&&\times\pin{k_2}\pin{k_3}\frac{\mb_{k_2}\mb_{k_3}}{\ok 0-\ok 2-\ok 3}2\pi\delta(k_0-k_2-k_3) |k_2 k_3\rangle_0.\nonumber
\eea
Notice that in either case, $k_0$ and $k_2+k_3$ differ by of order $1/\sigma$, which is by assumption much less than $m$.  Therefore $\ok{0}$ is quite far from $\ok{2}+\ok{3}$, and any creation of mesons of energies $\ok{2}$ and $\ok{3}$ from the initial wave packet will be far off-shell.  Thus we expect that such terms do not contribute to the meson multiplication probability.  Do they?

To answer this question, we need only calculate the reduced inner product of $|t\rangle$ with $|k_2k_3\rangle_\rv$ in Eq.~(\ref{sf}).

After the collision and to the order of $O(\sqrt{\lambda})$, $|k_2k_3\rangle_\rv$ is given in Eq.~(\ref{sf2}).
We can easily read the coefficients $\gamma$'s off from the states. We will always take the limit $\sigma\rightarrow\infty$ and first, let's consider the inner product after the collision.
\bea \label{gamma0k2k3}
\gamma_t^{01}(k)&=&\mb_{k}e^{-i\ok{} t}2\pi\delta(k-k_0)\\
\gamma_t^{02}(k\p_2k\p_3)&=&\frac{\sqrt{\lambda}V^{(3)}(\sqrt{\lambda}f(+\infty))\mb_{k_0}\md_{-k_0}\md_{k\p_2}\md_{k\p_3}e^{-i\ok{0}t}2\pi\delta(k_0-k\p_2-k\p_3)}{4\ok 0\left(\ok 0-\okp 2-\okp 3\right)}\nonumber\\
\gamma_{k_2k_3,\rv}^{02}(k\p_2k\p_3)&=&2\pi \delta(k\p_2-k_2)2\pi \delta(k\p_3-k_3)\nonumber\\
\gamma_{k_2k_3,\rv}^{01}(k)&=&\frac{\sl V^{(3)}(\sl f(+\infty))\md_{-k_2}\md_{-k_3}\md_{k}2\pi \delta(k-k_2-k_3)}{4\ok2 \ok3 (\ok2+\ok3-\ok{})}.\nonumber
\eea
Now we again use our master formula Eq.~(\ref{padqft}) to calculate the reduced inner product to $O(\sqrt{\lambda})$
\bea
{}_{\rv}\langle k_2k_3|t\rangle_{\rm{red}}&=&\ppin{k}\frac{\gamma_{k_2k_3,\rv}^{01*}(k)}{2 \ok{}}\sqrt{Q_0}\gamma_t^{01}(k)+2\int\hspace{-17pt}\sum \frac{dk\p_2dk\p_3}{(2\pi)^2}\frac{\gamma_{k_2k_3,\rv}^{02*}(k\p_2k\p_3)}{4\okp2\okp3}\sqrt{Q_0}\gamma_t^{02}(k\p_2k\p_3)\nonumber\\
&=&\ppin{k}\frac{\sqrt{Q_0}}{2 \ok{}}\frac{\sl V^{(3)}(\sl f(+\infty))\md_{k_2}\md_{k_3}\md_{-k}2\pi \delta(k-k_2-k_3)}{4\ok2 \ok3 (\ok2+\ok3-\ok{})}\mb_{k}e^{-i\ok{} t}2\pi\delta(k-k_0)\nonumber\\
&&+2\int\hspace{-17pt}\sum \frac{dk\p_2dk\p_3}{(2\pi)^2}\frac{\sqrt{Q_0}}{4\okp2\okp3}2\pi \delta(k\p_2-k_2)2\pi \delta(k\p_3-k_3)\nonumber\\
&&\times\frac{\sqrt{\lambda}V^{(3)}(\sqrt{\lambda}f(+\infty))\mb_{k_0}\md_{-k_0}\md_{k\p_2}\md_{k\p_3}e^{-i\ok{0}t}2\pi\delta(k_0-k\p_2-k\p_3)}{4\ok 0\left(\ok 0-\okp 2-\okp 3\right)}\nonumber\\
&=&\frac{\sqrt{\lambda Q_0} V^{(3)}(\sl f(+\infty))\mb_{k_2+k_3}\md_{k_2}\md_{k_3}\md_{-k_2-k_3}e^{-i\omega_{k_2+k_3} t}2\pi\delta(k_0-k_2-k_3)}{8\ok2 \ok3 \omega_{k_2+k_3}(\ok2+\ok3-\omega_{k_2+k_3})}\nonumber\\
&&+\frac{\sqrt{\lambda Q_0} V^{(3)}(\sl f(+\infty))\mb_{k_2+k_3}\md_{k_2}\md_{k_3}\md_{-k_2-k_3}e^{-i\omega_{k_2+k_3} t}2\pi\delta(k_0-k_2-k_3)}{8\ok2 \ok3 \omega_{k_2+k_3}(\omega_{k_2+k_3}-\ok2-\ok3)}\nonumber\\
&=&0.
\eea

The calculation of the inner product before the collision is similar. Here the $|k_2k_3\rangle_{\rm{vac}}$ that appear in the projector, and so in the matrix element, are given in Eq.~(\ref{sf}).  Therefore two of the $\gamma's$ in Eq.~(\ref{gamma0k2k3}) become
\bea
\gamma_t^{02}(k\p_2k\p_3)&=&\frac{\sqrt{\lambda}V^{(3)}(\sqrt{\lambda}f(-\infty))\mb_{k\p_2}\mb_{k\p_3}e^{-i\ok{0}t}2\pi\delta(k_0-k\p_2-k\p_3)}{4\ok 0\left(\ok 0-\okp 2-\okp 3\right)}\nonumber\\
\gamma_{k_2k_3,\rv}^{01}(k)&=&\frac{\sl V^{(3)}(\sl f(-\infty))\mb_{-k_2}\mb_{-k_3}\mb_{k}2\pi \delta(k-k_2-k_3)}{4\ok2 \ok3 (\ok2+\ok3-\ok{})}.\nonumber
\eea
Again to $O(\sl)$
\bea
{}_{\rv}\langle k_2k_3|t\rangle_{\rm{red}}&=&\ppin{k}\frac{\gamma_{k_2k_3,\rv}^{01*}(k)}{2 \ok{}}\sqrt{Q_0}\gamma_t^{01}(k)+2\int\hspace{-17pt}\sum \frac{dk\p_2dk\p_3}{(2\pi)^2}\frac{\gamma_{k_2k_3,\rv}^{02*}(k\p_2k\p_3)}{4\ok2\ok3}\sqrt{Q_0}\gamma_t^{02}(k\p_2k\p_3)\nonumber\\
&=&\ppin{k}\frac{\sqrt{Q_0}}{2 \ok{}}\frac{\sl V^{(3)}(\sl f(-\infty))\mb_{k_2}\mb_{k_3}\mb_{-k}2\pi \delta(k-k_2-k_3)}{4\ok2 \ok3 (\ok2+\ok3-\ok{})}\mb_{k}e^{-i\ok{} t}2\pi\delta(k-k_0)\nonumber\\
&&+2\int\hspace{-17pt}\sum \frac{dk\p_2dk\p_3}{(2\pi)^2}\frac{\sqrt{Q_0}}{4\ok2\ok3}2\pi \delta(k\p_2-k_2)2\pi \delta(k\p_3-k_3)\nonumber\\
&&\times\frac{\sqrt{\lambda}V^{(3)}(\sqrt{\lambda}f(-\infty))\mb_{k\p_2}\mb_{k\p_3}e^{-i\ok{0}t}2\pi\delta(k_0-k\p_2-k\p_3)}{4\ok 0\left(\ok 0-\okp 2-\okp 3\right)}\nonumber\\
&=&\frac{\sqrt{\lambda Q_0} V^{(3)}(\sl f(+\infty))\mb_{k_2}\mb_{k_3}e^{-i\omega_{k_2+k_3} t}2\pi\delta(k_0-k_2-k_3)}{8\ok2 \ok3 \omega_{k_2+k_3}(\ok2+\ok3-\omega_{k_2+k_3})}\nonumber\\
&&+\frac{\sqrt{\lambda Q_0} V^{(3)}(\sl f(-\infty))\mb_{k_2}\mb_{k_3}e^{-i\omega_{k_2+k_3} t}2\pi\delta(k_0-k_2-k_3)}{8\ok2 \ok3 \omega_{k_2+k_3}(\omega_{k_2+k_3}-\ok2-\ok3)}=0.
\eea
%\bea
%{}_{\rv}\langle k_2k_3|t\rangle&=&\frac{V^{(3)}(\sqrt{\lambda}f(-\infty))\sqrt{\lambda}e^{-i\ok{0}t}}{4\ok 0}\\
%&&\times\pin{k\p_2}\pin{k\p_3}\frac{\mb_{k\p_2}\mb_{k\p_3}}{\ok 0-\okp 2-\okp 3}2\pi\delta(k_0-k\p_2-k\p_3) \ {}_0\langle k_2k_3|k\p_2 k\p_3\rangle_0\nonumber\\
%&&+\frac{\sl V^{(3)}(\sl f(-\infty))\mb_{k_2}\mb_{k_3}\mb_{-k_2-k_3}}{4\ok2\ok3(\ok2+\ok3-\omega_{k_2+k_3})}\nonumber\\
%&&\times 2\sigma\sqrt{\pi}\mb_{k_0}e^{-i\ok{0}t}\pin{k_1}e^{-\sigma^2\left(k_1-k_0\right)^2}e^{-i(k_1-k_0)\left(x_0+\frac{k_0}{\ok{0}}t\right)}{}_0\langle k_2+k_3|k_1\rangle_0 \nonumber\\
%%&=&\frac{V^{(3)}(\sqrt{\lambda}f(-\infty))\sqrt{\lambda}e^{-i\ok{0}t}}{4\ok 0}\nonumber\\
%%&&\times\pin{k\p_2}\pin{k\p_3}\frac{\mb_{k\p_2}\mb_{k\p_3}}{\ok 0-\okp 2-\okp 3}2\pi\delta(k_0-k\p_2-k\p_3)\nonumber\\
%%&&\times \frac{{}_0{\langle 0}|0\rangle_0}{4 \omega_{k_2} \omega_{k_3}} \left(2 \pi \delta\left(k_2^{\prime}-k_2\right) 2 \pi \delta\left(k_3^{\prime}-k_3\right)+2 \pi \delta\left(k_2^{\prime}-k_3\right) 2 \pi \delta\left(k_3^{\prime}-k_2\right)\right)\nonumber\\
%&=&\frac{V^{(3)}(\sqrt{\lambda}f(-\infty))\sqrt{\lambda}\mb_{k_2}\mb_{k_3}e^{-i\ok{0}t}2\pi\delta(k_0-k_2-k_3)\ {}_0{\langle 0}|0\rangle_0}{8\ok 0\ok 2\ok 3\left(\ok 0-\ok 2-\ok 3\right)}\nonumber\\
%&&+\frac{\sl V^{(3)}(\sl f(-\infty))\mb_{k_2}\mb_{k_3}e^{-i\ok{0}t}2\pi\delta(k_0-k_2-k_3)\ {}_0{\langle 0}|0\rangle_0}{8\ok0\ok2\ok3(\ok2+\ok3-\ok0)}=0.\nonumber
%\eea
We see that the inner product also vanishes.  Thus initial and final state corrections due not contribute at this order away from the pole.  Of course this is to be expected, as the process only conserves energy at the pole.

\subsection{The Pole Contribution} \label{polesez}

In Eq.~(\ref{speccon}) we calculated the contribution of the term (\ref{spec}) to the meson multiplication amplitude, and found that it vanished.  However we ignored the contribution from the pole at $\ok 1=\ok 2+\ok 3$.  More precisely, we set $k_1$ to $k_0$ in the denominator, although in general they differ by of order $O(1/\sigma)$.  This approximation is reasonable except in a neighborhood of size $1/\sigma$ of the pole.  In the limit $\sigma\rightarrow\infty$ this becomes valid except in an infinitesimal neighborhood of the pole.  One thus expects that the error introduced depends only on the integrand in that infinitesimal neighborhood, and in particular only on the residue of the pole.  %We will now modify the integral, preserving the residue.  As we have just argued, only the residue can contribute to the amplitude and so this modification will not affect the amplitude.

At this pole, energy is conserved, and so this contribution to the multiplication would be on-shell.  Let us rewrite the contribution (\ref{speccon}) to the amplitude, now keeping the contribution from the pole
\bea
|t\rangle&\supset&\frac{\sqrt{\lambda}\mb_{k_0}e^{-i\ok{0}t}}{4}\pin{k_2}\pin{k_3}\int dx \V3 \g_{k_2}(x)\g_{k_3}(x)\nonumber\\
&&\times  2\sigma\sqrt{\pi}\left[\pin{k_1} \frac{e^{-\sigma^2\left(k_1-k_0\right)^2}e^{-i(k_1-k_0)x_t}}{\ok 1-\ok 2-\ok 3}\frac{\g_{-k_1}(x)}{\ok 1}\right]|k_2 k_3\rangle_0. \label{dint}
\eea
As is written, the integral is not defined at the pole.  Let us, for now, assert that it is defined using a principal value prescription.

Let us define the location of the pole by $k_1=k_I$ such that
\beq
\ok I=\ok 2+\ok 3\hsp k_I>0.
\eeq
There is another pole at $k_1=-k_I$.  We can evaluate the contributions to the term in brackets at the poles using the Sokhotski–Plemelj theorem.  We have already argued that the contributions away from the poles do not contribute to the amplitude, so we only need to consider $\pm i\pi$ times the residue at each pole.  At the pole $k_1=-k_I$ the residue contains a factor of $e^{-\sigma^2(k_I+k_0)^2}$ which vanishes in the limit $\sigma\rightarrow\infty$ and so we will not consider that pole further.

%but this is irrelevant in the limit $\sigma\rightarrow\infty$ as a result of the Gaussian factor.  We have shown in Subsec.~\ref{isc} that only the contributions to $|t\rangle$ at the poles themselves can contribute to the amplitude.  This means that the amplitude only depends on the residues at the two poles, and the integrand may be changed away from the poles, changing the integral but leaving the amplitude invariant. At the pole $k_1=k_I$ we replace $k_1$ with $k_I$ in the Gaussian factor, as this does not affect the residue at $k_1=k_I$, while at $k_1=-k_I$ we replace it with $-k_I$, preserving the fact that  the corresponding residue vanishes at $\sigma\rightarrow\infty$.  We won't consider this pole further.

If $x_t>x$ then the contour should be closed below yielding $-\pi i$ times the residue, otherwise it should be closed above yielding $\pi i$ times the residue.  Note that naively the Gaussian term diverges on such a contour.  However, it only contributes a constant factor to the integral in a $1/\sigma$-neighborhood of the pole in the $\sigma\rightarrow\infty$ limit, and so one can simply set the $k_1$ in the Gaussian to its value at the pole before performing the integration.  This affects the value of the integral over the real line, but as we have argued, only the integral in a neighborhood of the pole can contribute to the amplitude.  The term in brackets in (\ref{dint}) thus becomes%\footnote{A more direct and contour-free approach is as follows.  The odd part of the $e^{-i(k_1-k_0)(x_t-x)}$ and the pole combine to form a sinc function, which in the limit $\sigma\rightarrow\infty$ becomes a Dirac delta function times sign$(x_t-x)$.  Using the delta function to perform the $k_1$ integral one arrives at the same result.}
\beq
-\sign{x_t-x}\frac{i}{2k_I}  e^{-\sigma^2\left(k_I-k_0\right)^2} e^{-i(k_I-k_0)x_t}\g_{-k_I}(x). \label{fint}
\eeq

An alternate derivation, without use of contour integrals, is as follows.  At large $|x|$ the $\g_{-k_1}(x)$ in the square bracket, up to a constant phase $\mb_{-k_1}$ or $\md_{-k_1}$, is just $e^{i k_1 x}$.  Combining this with the $e^{-i(k_1-k_0)x_t}$ yields
\beq
e^{-i(k_1-k_0)x_t}e^{i k_1 x}
=e^{-i(k_1-k_I)(x_t-x)}e^{-i(k_I-k_0)x_t}e^{ik_Ix}. \label{fasa}
\eeq
The third term on the right hand side, together with the phase $\mb_{-k_1}$ or $\md_{-k_1}$, becomes the $g_{-k_I}(x)$ in (\ref{fint}).   The second also appears in (\ref{fint}).  At large $\sigma$, we may expand the denominator $(\ok1-\ok I)\ok 1$ of the term in brackets to linear order in $(k_1-k_I)$, yielding $(k_1-k_I)k_1$.  This denominator is odd in $(k_1-k_I)$, and so only the odd term in the first term on the right hand side of (\ref{fasa}) contributes.  Dividing this by $(k_1-k_I)k_1$ one identifies the nascent delta function
\beq
\lim{|x_t-x|\rightarrow\infty}-\frac{{\rm sin}\left[(k_1-k_I)(x_t-x)  \right]}{(k_1-k_I)k_1}=-\pi{\rm sign}(x_t-x)\frac{\delta(k_1-k_I)}{k_1}
\eeq
which can then be used to perform the $k_1$ integral in the square brackets in Eq.~(\ref{dint}), leading again to Eq.~(\ref{fint}).

This is not quite what we expect.  It is, up to a sign, symmetric in time and so the probability of observing two mesons will be the same in the far past and the far future.

Let us break the time reversal symmetry by changing our prescription for interpreting the pole in $\gamma_{1\kt}^{02}$.  Instead of the principal value prescription, let us try
\beq
\gamma_{1\kt}^{02}(k_1,k_2)= \frac{2\pi\delta(k_2-\kt)}{2}\left(-\Delta_{k_1 B}-\sqrt{Q_0\lambda}\frac{V_{\I  k_1}}{\ok1}\right)+\frac{\sqrt{Q_0\lambda}V_{-\kt k_1 k_2}}{4\omega_\kt\left(\omega_\kt-\ok1-\ok2+i\epsilon\right)}. \label{newinit}
\eeq
In the next subsection we will explain why such a shift leads to another Hamiltonian eigenstate with the same energy and so is allowed.

Now the pole on the complex $k_1$ plane is at an infinitesimal negative imaginary value.  As a result, if $x_t<x$ then the pole is not included in the contour.  Now the term in brackets becomes
\beq
-\Theta(x_t-x)\frac{i}{k_I}  e^{-\sigma^2\left(k_I-k_0\right)^2} e^{-i(k_I-k_0)x_t}\g_{-k_I}(x)
\eeq
where $\Theta$ is the Heaviside step function.  The corresponding contribution to $|t\rangle$ is
\bea
|t\rangle&\supset&-\frac{\sqrt{\lambda}\mb_{k_0}e^{-i\ok{0}t}}{4}\pin{k_2}\pin{k_3}\int_{-\infty}^{x_t} dx \V3 \g_{k_2}(x)\g_{k_3}(x)\nonumber\\
&&\times  2\sigma\sqrt{\pi}\left[\frac{i }{k_I}  e^{-\sigma^2\left(k_I-k_0\right)^2} e^{-i(k_I-k_0)x_t}\g_{-k_I}(x)\right]|k_2 k_3\rangle_0.
\eea
Now if $x_t\ll 0$, so that the meson wave packet has not reached the kink, then the $x$-integral will only cover the asymptotic region where $\g_{k_2}(x)\g_{k_3}(x)\g_{-k_I}(x)\sim e^{ix(k_I-k_2-k_3)}$ oscillates rapidly, exponentially suppressing the amplitude.  On the other hand, after the collision $x_t\gg 0$ and so the integral is, up to an exponentially suppressed correction, equal to $V_{-k_Ik_2k_3}$.  The state is then
\beq
|t\rangle\supset-\Theta(x_t)\frac{i\sigma\sqrt{\pi\lambda}\mb_{k_0}e^{-i\ok{0}t}}{2 }\pin{k_2}\pin{k_3} e^{-\sigma^2\left(k_I-k_0\right)^2} e^{-i(k_I-k_0)x_t}\frac{V_{-k_I k_2 k_3}}{k_I}|k_2 k_3\rangle_0.
\eeq
%This leads to the reduced matrix-element
%\beq
%\frac{\langle k_2 k_3|t\rangle}{\langle 0\vac}\supset\sign{x_t}\frac{i\sigma\sqrt{\pi\lambda}\mb_{-k_0}e^{-i\ok{0}t}}{8\ok 2\ok 3k_I }e^{-\sigma^2\left(k_I-k_0\right)^2} e^{-i(k_I-k_0)x_t}V_{-k_I k_2 k_3}
%\eeq
%plus corrections of order $O(\lambda^{3/2})$.

Using (\ref{grred}) to evaluate the denominator, this leads to the reduced matrix-element
\beq
\frac{{{}_{\rm{vac}}\langle k_2 k_3|t\rangle_{\rm{red}}}}{{}_{\rm{}}\langle 0|0\rangle_{\rm{red}}}\supset-\Theta(x_t)\frac{i\sigma\sqrt{\pi\lambda}\mb_{k_0}e^{-i\ok{0}t}}{4\ok 2\ok 3k_I }e^{-\sigma^2\left(k_I-k_0\right)^2} e^{-i(k_I-k_0)x_t}V_{-k_I k_2 k_3}
\eeq
plus corrections of order $O(\lambda^{3/2})$, in agreement with the amplitude reported in Ref.~\cite{memult}, noting that in the large $\sigma$ limit the $e^{-i(k_I-k_0)x_t}$ term tends to $1$.  However, here we have considered a translation-invariant initial and final state.  Thus we conclude that the higher order corrections to the initial and final state which lead to translation-invariance do not affect the meson multiplication amplitude at order $O(\sqrt{\lambda}).$

%Note that $\sign{x_t}$ is equal to $-1$ before the meson wave packet reaches the kink, and $+1$ after.  Thus one observes two-meson states both before and after the collision.  However the amplitude if one-half of that reported in Ref.~\cite{memult}.

%Needless to say, the resolution to both of these puzzles is clear.  The initial state in our experiment has only one meson, and so does not correspond to the state $|\Phi\rangle$ defined above.  Shifting the 2-meson part of the wave function to zero, before the collision, will double the amplitude after the collision, restoring agreement with Ref.~\cite{memult}.  Let us now explain why such a shift is justified.

\subsection{Degenerate Eigenstates}

%We have found that no contributions to the states $|k_1\rangle$ given in Eq.~(\ref{gammakt}) contribute to the meson multiplication amplitude at order $O(\sqrt{\lambda})$.  Are these the only contributions  to the initial states?

We defined $|k_1\rangle$ to be the $H\p$ eigenstate which is annihilated by $P\p$ and whose leading order term is $|k_1\rangle_0$.  This does not completely characterize the state, because there are other translation-invariant states with the same energy.  Consider any $k_2$ and $k_3$ such that $\tilde{\omega}_{k_2}+\tilde{\omega}_{k_3}=\tilde{\omega}_{k_1}$.  Recall that, up to corrections of order $O(\lambda)$, which we do not consider, this condition is $\ok{2}+\ok{3}=\ok{1}$.  Then the state $|k_2 k_3\rangle$ has the same energy as $|k_1\rangle$ and it is, by construction, also translation invariant.  

Let us shift the definition of $|k_1\rangle$ by
\beq
|k_1\rangle\longrightarrow |k_1\rangle+c_{k_1k_2k_3}\sqrt{\lambda}|k_2 k_3\rangle
\eeq
where $c$ is of order $O(\lambda^0)$ and is nonvanishing only when $\tilde{\omega}_{k_2}+\tilde{\omega}_{k_3}=\tilde{\omega}_{k_1}$.  Now the energy eigenvalues match in the new term, so the argument used above, to argue that the contributions to $|k_1\rangle$ in $\gamma_{1k_1}^{m2}$ do not contribute to the amplitude, cannot be applied.  

This new choice of $|k_1\rangle$ also satisfies our definition.   However it differs from the old choice by a change in $\gamma_1^{20}(k_2,k_3)$.  In fact, any value of $\gamma_1^{20}(k_2,k_3)$ corresponds to some choice of $c_{k_1k_2k_3}$ so long as it agrees with the old value in Eq.~(\ref{gammakt}) when $\ok 1\neq \ok 2+\ok 3$.  Intuitively, one may only add something proportional to $\delta(\ok 1-\ok 2-\ok 3)$.  The infinitesimal shift in the pole in (\ref{newinit}) is exactly of this form.

%What value of $c_{k_1k_2k_3}$ can fix the two problems identified in the previous subsection?  Recall that at $x_t<0$ our contour ran through the middle of the pole and closed on the bottom of the complex plane, yielding $-\pi i$, when we needed $0$ to eliminate the 2-meson part of the initial state.  Similarly at $x_t>0$ the contour ran through the middle of the pole and closed on the top of the complex plane, yielding $\pi i$ when we needed $2\pi i$ to recover the amplitude from Ref.~\cite{memult}.  

%Both of these problems would be cured if the contour ran beneath the pole, and so if the pole is shifted by $i\epsilon$.  This can be arranged, without changing $\gamma_1^{20}(k_2,k_3)$ away from the pole, if one chooses $c$ to shift it to
%\beq
%\gamma_{1\kt}^{02}(k_1,k_2)= \frac{2\pi\delta(k_2-\kt)}{2}\left(\Delta_{k_1 B}-\sqrt{Q_0\lambda}\frac{V_{\I  k_1}}{\ok1}\right)+\frac{\sqrt{Q_0\lambda}V_{-\kt k_1 k_2}}{4\omega_\kt\left(\omega_\kt-\ok1-\ok2-i\epsilon\right)}. \label{newinit}
%\eeq

We thus claim that the correct initial condition in Ref.~\cite{memult} corresponds to Eq.~(\ref{gammakt}) with $\gamma_1^{21}$ replaced by Eq.~(\ref{newinit}).   What if the meson wave packet scatters with the kink from the other side?  Then $x_0>0$ and $k_0<0$.  In this case, we want the integral to vanish when $x_t<x$, so that the $k_1$ contour is closed on the bottom of the complex plane.  This requires the pole to be shifted by $+i\epsilon$.  However, as $k_0<0$, this still corresponds to a negative imaginary part for $\ok 1$, and so still corresponds to the modification (\ref{newinit}).   We remind the reader that this state has the same energy, momentum and $O(\lambda^0)$ term as the state defined by Ref.~\cite{memult}, but does not lead to a two-meson component in the initial wave packet $|\Phi\rangle$.

\section{Remarks}

Given a stationary kink solution, one may compute its normal modes and even their interactions \cite{shapeinter}.  Every year, this is done for new classes of models \cite{wshifman,takyi1,takyi2}, including recently even gravitating kinks \cite{yuan1,yuan2}.  With these normal modes in hand, one can construct the quantum states corresponding to kinks.   Recently there has even been progress towards to a quantum treatment of nontopological solitons \cite{quantosc,kovbreather}.  However every such treatment needs to deal with the fact that translation-invariant kink states are non-normalizable, as a result of the infinite volume of the translation group.

There are many proposed solutions to this problem, each useful in some settings.  Many have the drawback that they destroy translation-invariance, they do not preserve local quantities or they cause finite shifts.  In the present note, we have proposed another method of dealing with this problem, replacing inner products by reduced inner products where we have quotiented by the translation group.  This is, in our opinion, a reasonable approach as the volume of the translation group appears in the numerator and denominator of observable quantities and so is canceled.  We have found that this greatly simplifies many calculations, as we are able to fix the translation symmetry so that all terms with zero modes $\phi_0$ vanish.  

The problem was complicated by our choice of coordinates $y$, defined to be the eigenvalue of $\phi_0$.  Intuitively, this can be understood as follows.  Let $f(x)$ be a classical kink solution.  A shift in the collective coordinate transforms $f(x)$ to $f(x-x_0)$, and so acts linearly on the position.  On the other hand, a shift in $y$ changes $f(x)$ to $f(x)-y_0 f\p(x_0)/\sqrt{Q_0}$.  This does not correspond to a shifted kink solution, unless one simultaneously compensates by shifting the normal modes.  Thus the nondiagonal part of the Jacobian factor arising from a quotient by this translation symmetry is proportional to $\Delta_{Bk}$, which is the mixing between the zero mode mode $\g_B(x)$ and the other normal modes, when one shifts $x$.  However, at small $y_0$ these two transformations are related by a simple proportionality factor of $\sqrt{Q_0}$, which allows us to easily define a matching condition and evaluate the necessary Jacobian.

In Ref.~\cite{menormal}, the leading corrections to a 1-meson state $|\kt\rangle$ were found, summarized here in Eq.~(\ref{gammakt}).  However, if $\omega_{\kt}\geq 2m$ then this state has the same energy and momenta as some 2-meson states.  The state always contains a cloud of off-shell 2-meson states.  In Eq.~(\ref{gammakt}), the degenerate states are on-shell.  The physically correct initial condition to study 2-meson production is to begin with a state that does not contain a component with two on-shell mesons.   In Eq.~(\ref{newinit}) we present this state.  It is equal to that of Ref.~\cite{menormal}, with a subleading contribution from a degenerate 2-meson state.  This subleading contribution is added by including an infinitesimal, imaginary shift of a pole.   We suspect more generally that such poles in states represent on-shell contributions from degenerate states, which can and often should be removed via such imaginary shifts.  Said differently, we suspect that the prescription for evaluating such poles corresponds in general to a physical choice of Hamiltonian eigenstate in a degenerate eigenspace.

%The contribution of the $c$ term to the state at time $t$ is
%\bea
%|t\rangle&\supset&\sqrt{\lambda}\mb_{-k_0}e^{-i\ok{0}t}\pin{k_2}\pin{k_3}%\frac{1}{\ok 0-\ok 2-\ok 3}\int dx \V3 \g_{k_2}(x)\g_{k_3}(x)
%\nonumber\\
%&&\times  
%\left[2\sigma\sqrt{\pi}\pin{k_1} e^{-\sigma^2\left(k_1-k_0\right)^2}e^{-i(k_1-k_0)\left(x_0+\frac{k_0}{\ok{0}}t\right)}c_{k_1k_2k_3}\right]|k_2 k_3\rangle_0
%\eea

%Now, the subleading terms in the projector which eliminate the $|k_2k_3\rangle$ terms in the initial state (\ref{prima}), will subtract just the same terms from the final state (\ref{dopo}).  Thus they will leave the difference between the two
%\bea
%\Delta|t\rangle&=&\frac{2\sigma\sqrt{\pi}\left(V^{(3)}(\sqrt{\lambda}f(-\infty))-V^{(3)}(\sqrt{\lambda}f(\infty))\right)\sqrt{\lambda}\mb_{-k_0}\g_{-k_0}(x_t)e^{-i\ok{0}t}}{4\ok 1}\\
%&&\times\pin{k_2}\pin{k_3}\frac{\g_{k_2}(x_t)\g_{k_3}(x_t)}{\ok 0-\ok 2-\ok 3}e^{-\sigma^2(k_0-k_2-k_3)^2} |k_2 k_3\rangle_0\nonumber
%\eea

%\alpha_{k_1}=2\sigma\sqrt{\pi}\mb_{-k_1}e^{-\sigma^2\left(k_1-k_0\right)^2}e^{i(k_0-k_1)x_0}.

%\frac{\sqrt{Q_0\lambda}V_{-\kt k_1 k_2}}{4\omega_\kt\left(\omega_\kt-\ok1-\ok2\right)}

\section* {Acknowledgement}

\noindent
JE is supported by NSFC MianShang grants 11875296 and 11675223. HL acknowledges the support from CAS-DAAD Joint Fellowship Programme for Doctoral students of UCAS.

\end{document}

\section{Stokes Scattering}

\bea
H_I&=&\frac{\sqrt{\lambda}}{2} \int \frac{d k_1}{2 \pi} \frac{d k_2}{2 \pi}  \frac{V_{-k_1 k_2 S}}{\omega_{k_1}} B_{k_2}^{\ddagger} B_{S}^{\ddagger} B_{k_1} \\
V_{-k_1 k_2 S}&=&\int d x V^{(3)}(\sqrt{\lambda} f(x)) \mathfrak{g}_{-k_1}(x) \mathfrak{g}_{k_2}(x) \mathfrak{g}_{S}(x).\nonumber
\eea

\beq
\Phi(x)=\operatorname{Exp}\left[-\frac{\left(x-x_0\right)^2}{4 \sigma^2}+i x k_0\right], \quad x_0 \ll-\frac{1}{ m}, \quad  \frac{1}{k_0},\frac{1}{m}\ll\sigma \ll\left|x_0\right| .
\eeq

\begin{equation}
\alpha_k=\int d x \Phi(x) \mathfrak{g}_k^*(x).
\end{equation}

\beq
|S k\rangle=B_s^\ddag B^\ddag_k\vac_0\hsp |k\rangle=B^\ddag_k\vac_0.
\eeq

\begin{equation}
\left|\Phi\right\rangle=\pin{k_1} \alpha_{k_1}\left|k_1\right\rangle
\end{equation}

\beq
e^{-it(H\p_2+H_I)}=e^{-itH\p_2}-i\int _0^t dt_1 e^{-i(t-t_1)H\p_2}H_I e^{-i t_1 H\p_2}+O(\lambda)
\eeq

\beq
\ok{I}=\ok{2}+\os\hsp k_I>0
\eeq

\beq
e^{-iH t}|k_1\rangle=\frac{-i\sqrt{\lambda}}{2\omega_{k_1}} \pin{k_2}V_{S,k_2,-k_1}e^{-\frac{it}{2}(\omega_{k_1}+\os+\ok{2})}\frac{\rm{sin}\left[\left(\frac{\os+\ok{2}-\ok{1}}{2}\right)t\right]}{(\os+\ok{2}-\ok{1})/2}|S k_2\rangle
\eeq

\beq
\frac{\rm{sin}\left[\left(\frac{\os+\ok{2}-\ok{1}}{2}\right)t\right]}{(\os+\ok{2}-\ok{1})/2}=2\pi\delta(\os+\ok{2}-\ok{1})=\left(\frac{\ok{I}}{k_I}\right)\left(2\pi\delta(k_1-k_I)+2\pi\delta(k_1+k_I)\right)
\eeq

\bea
e^{-iH t}|\Phi\rangle&=&-i\sqrt{\lambda}\pin{k_1}\frac{\alpha_{k_1}}{2\ok{1}} \pin{k_2}V_{S,k_2,-k_1}e^{-\frac{it}{2}(\ok{1}+\os+\ok{2})}\frac{\rm{sin}\left[\left(\frac{\os+\ok{2}-\ok{1}}{2}\right)t\right]}{(\os+\ok{2}-\ok{1})/2}|S k_2\rangle\nonumber\\
&=&\frac{-i\sqrt{\lambda}}{2} \pin{k_2}e^{-i\ok{I}t}\left(\frac{1}{k_I}\right)\left(\alpha_{k_I}V_{S,k_2,-k_I}+\alpha_{-k_I}V_{S,k_2,k_I}
\right)|S k_2\rangle
\eea

\bea
\g_k(x)&=&\left\{\begin{tabular}{lll}
$\mb_ke^{ikx}+\mc_ke^{-ikx}$&\rm{if} & $x\ll  -1/m$\\
$\md_ke^{ikx}+\me_k e^{-ikx}$&\rm{if} & $x\gg 1/m$\\
\end{tabular}
\right. \label{gk}\\
|\mb_k|^2+|\mc_k|^2&=&|\md_k|^2+|\me_k|^2=1\hsp
\mb^*_k=\mb_{-k}\hsp
\mc^*_k=\mc_{-k}\hsp
\md^*_k=\md_{-k}\hsp
\me^*_k=\me_{-k}.\nonumber
\eea

\bea
\alpha_{k_I}&=&2\sigma\sqrt{\pi}\left[\mb^*_{k_I}e^{ix_0(k_0-k_I)}e^{-\sigma^2(k_0-k_I)^2}+\mc^*_{k_I}e^{ix_0(k_0+k_I)}e^{-\sigma^2(k_0+k_I)^2}
\right]\\
&=&2\sigma\sqrt{\pi}\mb^*_{k_I}e^{ix_0(k_0-k_I)}e^{-\sigma^2(k_0-k_I)^2}\nonumber
\eea

\bea
\alpha_{-k_I}&=&2\sigma\sqrt{\pi}\left[\mb_{k_I}e^{ix_0(k_0+k_I)}e^{-\sigma^2(k_0+k_I)^2}+\mc_{k_I}e^{ix_0(k_0-k_I)}e^{-\sigma^2(k_0-k_I)^2}
\right]\\
&=&2\sigma\sqrt{\pi}\mc_{k_I}e^{ix_0(k_0-k_I)}e^{-\sigma^2(k_0-k_I)^2}\nonumber
\eea

\bea
e^{-iH t}|\Phi\rangle&=&-i\sigma\sqrt{\pi\lambda} \pin{k_2}e^{ix_0(k_0-k_I)}e^{-\sigma^2(k_0-k_I)^2}e^{-i\ok{I}t}\left(\frac{\tilde{V}_{S,k_2,-k_I}}{k_I}\right)|S k\rangle\\
\tilde{V}_{S,k_2,-k_I}&=&\mb^*_{k_I}V_{S,k_2,-k_I}+\mc_{k_I}V_{S,k_2,k_I}
\nonumber
\eea

\beq
\langle S k_1|S k_2\rangle=\frac{2\pi\delta(k_1-k_2)}{4\os\ok{1}}{}_0\langle 0\vac_0.
\eeq

\beq
\frac{\langle S k_2|e^{-iH t}|\Phi\rangle}{{}_0\langle 0\vac_0}=\frac{-i\sigma\sqrt{\pi\lambda}}{4\os\ok{2}k_I} e^{ix_0(k_0-k_I)}e^{-\sigma^2(k_0-k_I)^2}e^{-i\ok{I}t}\tilde{V}_{S,k_2,-k_I}
\eeq

\bea
\left|\frac{\langle S k_2|e^{-iH t}|\Phi\rangle}{{}_0\langle 0\vac_0}\right|^2&=&\frac{\sigma^2\pi\lambda}{16\os^2\ok{2}^2k^2_I}\left|\tilde{V}_{S,k_2,-k_I}\right|^2e^{-2\sigma^2(k_0-k_I)^2}\\
&=&\frac{\sigma\pi^{3/2}\lambda}{16\sqrt{2}\os^2\ok{2}^2k^2_I}\left|\tilde{V}_{S,k_2,-k_I}\right|^2\delta(k_I-k_0)\nonumber
\eea

\beq
\mathcal{P}=\pin{k_2} \frac{4\os\ok{2}}{{}_0\langle 0\vac_0} |Sk_2\rangle\langle Sk_2|
\eeq

\beq
\frac{\langle k_1|k_2\rangle}{{}_0\langle 0\vac_0}=\frac{2\pi\delta(k_1-k_2)}{2\ok{1}}
\eeq

\bea
\frac{\langle\Phi|\Phi\rangle}{{}_0\langle 0\vac_0}&=&\pink{2}\alpha_{k_1}\alpha^*_{k_2}\frac{\langle k_2|k_1\rangle}{{}_0\langle 0\vac_0}=\pin{k}\frac{|\alpha_k|^2}{2\ok{}}=\frac{1}{2\omega_{k_0}}\pin{k}|\alpha_k|^2\\
&=&\frac{1}{2\omega_{k_0}}\pin{k}\int dx\int dy g_k^*(x)g_k(y)\Phi(x)\Phi^*(y)
\nonumber\\
&=&\frac{1}{2\omega_{k_0}}\int dx |\Phi(x)|^2=\frac{\sigma\sqrt{\pi}}{\sqrt{2}\omega_{k_0}}\nonumber
\eea

\bea
P&=&\frac{\langle\Phi|e^{iHt}\mathcal{P}e^{-iHt}|\Phi\rangle}{\langle\Phi|\Phi\rangle}=\pin{k_2} \frac{4\os\ok{2}}{{}_0\langle 0\vac_0}\frac{\left|\langle Sk_2|e^{-iHt}|\Phi\rangle\right|^2}{\langle\Phi|\Phi\rangle/{}_0\langle 0\vac_0}\frac{1}{{}_0\langle 0\vac_0}\\
&=&\pin{k_2}4\os\ok{2}\frac{\frac{\sigma\pi^{3/2}\lambda}{16\sqrt{2}\os^2\ok{2}^2k^2_I}\left|\tilde{V}_{S,k_2,-k_I}\right|^2\delta(k_I-k_0)}{\left(\frac{\sigma\sqrt{\pi}}{\sqrt{2}\omega_{k_0}}\right)}
\nonumber\\
&=&\frac{\pi\lambda\ok{0}}{4\os (\ok{0}-\os)k_0^2}\pin{k_2}\left|\tilde{V}_{S,k_2,-k_I}\right|^2\delta(k_I-k_0)\nonumber\\
&=&\lambda\frac{\left|\tilde{V}_{S,\sqrt{(\ok{0}-\ok{S})^2-m^2},-k_0}\right|^2+\left|\tilde{V}_{S,-\sqrt{(\ok{0}-\ok{S})^2-m^2},-k_0}\right|^2
}{8\os k_0\sqrt{(\ok{0}-\ok{S})^2-m^2}}\nonumber
\eea

\section{Anti-Stokes Scattering}

\bea
H_I&=&\frac{\sqrt{\lambda}}{4\os} \int \frac{d k_1}{2 \pi} \frac{d k_2}{2 \pi}  \frac{V_{-k_1 k_2 S}}{\omega_{k_1}} B_{k_2}^{\ddagger} B_{S} B_{k_1} %\\
%V_{-k_1 k_2 S}&=&\int d x V^{(3)}(\sqrt{\lambda} f(x)) \mathfrak{g}_{-k_1}(x) \mathfrak{g}_{k_2}(x) \mathfrak{g}_{S}(x).\nonumber
\eea

\beq
\left|\Phi\right\rangle=\pin{k_1} \alpha_{k_1}\left|S k_1\right\rangle
\eeq

\section{Example: $\phi^4$ Double-Well Model}

{\blu{Below I took the complex conjugate of the $\phi^4$ paper ref, is that the right way to change the convention?  $k\rightarrow -k$ wouldn't do anything}}
\beq
V_{k_1k_2S}=i\pi \frac{3\sqrt{3\lambda}}{8}\frac{\left(17\b^4-(\ok1^2-\ok2^2)^2\right)(\b^2+k_1^2+k_2^2)+8\b^2k_1^2k_2^2}{\b^{3/2}\ok1\ok2\sqrt{\b^2+k_1^2}\sqrt{\b^2+k_2^2}}\sech\left(\frac{\pi(k_1+k_2)}{2\b}\right).
\eeq

\section{Remarks}

\end{document}